\documentclass[%
 reprint,
superscriptaddress,
 amsmath,amssymb,
 aps,
]{revtex4-2}

\usepackage{graphicx}
\usepackage{dcolumn}
\usepackage{bm}

\begin{document}


\title{Exploring low-energy neutrino physics with the\\ Coherent Neutrino Nucleus Interaction Experiment (CONNIE)
}

\author{Alexis Aguilar-Arevalo}
\affiliation{Universidad Nacional Aut\'onoma de M\'exico, CDMX, M\'exico}

\author{Xavier Bertou}
\affiliation
{Centro At\'omico Bariloche and Instituto Balseiro, Comisi\'on Nacional de Energ\'ia At\'omica (CNEA), Consejo Nacional de Investigaciones Cient\'ificas y T\'ecnicas (CONICET), Universidad Nacional de Cuyo (UNCUYO), San Carlos de Bariloche, Argentina. }

\author{Carla Bonifazi}
\affiliation{Universidade Federal do Rio de Janeiro, Instituto de F\'isica, Rio de Janeiro, RJ, Brazil}

\author{Gustavo Cancelo}
\affiliation{Fermi National Accelerator Laboratory, Batavia, IL, United States}

\author{Alejandro Casta\~neda}
\affiliation{Universidad Nacional Aut\'onoma de M\'exico, CDMX, M\'exico}

\author{Brenda Cervantes Vergara}
\affiliation{Universidad Nacional Aut\'onoma de M\'exico, CDMX, M\'exico}

\author{Claudio Chavez}
\affiliation{Facultad de Ingenier\'ia - Universidad Nacional de Asunci\'on, Asunci\'on, Paraguay}

\author{Juan C. D'Olivo}
\affiliation{Universidad Nacional Aut\'onoma de M\'exico, CDMX, M\'exico}

\author{Jo\~ao C. dos Anjos}
\affiliation{Centro Brasileiro de Pesquisas F\'isicas, Rio de Janeiro, RJ, Brazil}

\author{Juan Estrada}
\affiliation{Fermi National Accelerator Laboratory, Batavia, IL, United States}

\author{Aldo R. Fernandes Neto}
\affiliation{Centro Federal de Educa\c c\~ao Tecnol\'ogica Celso Suckow da Fonseca, Angra dos Reis, RJ, Brazil}

\author{Guillermo Fernandez Moroni}
\affiliation{Fermi National Accelerator Laboratory, Batavia, IL, United States}
\affiliation{Instituto de Investigaciones en Ingenier\'ia El\'ectrica, Departamento de Ingenier\'ia El\'ectrica y Computadoras, Universidad Nacional del Sur (UNS) - CONICET, Bah\'ia Blanca, Argentina}

\author{Ana Foguel}
\affiliation{Universidade Federal do Rio de Janeiro, Instituto de F\'isica, Rio de Janeiro, RJ, Brazil}

\author{Richard Ford}
\affiliation{Fermi National Accelerator Laboratory, Batavia, IL, United States}

\author{Juan Gonzalez Cuevas}
\affiliation{Facultad de Ingenier\'ia - Universidad Nacional de Asunci\'on, Asunci\'on, Paraguay}

\author{Pamela Hern\'andez}
\affiliation{Universidad Nacional Aut\'onoma de M\'exico, CDMX, M\'exico}

\author{Susana Hernandez}
\affiliation{Fermi National Accelerator Laboratory, Batavia, IL, United States}

\author{Federico Izraelevitch}
\affiliation{Universidad Nacional de San Mart\'in (UNSAM), Comisi\'on Nacional de Energ\'ia At\'omica (CNEA),  Consejo Nacional de Investigaciones Cient\'ificas y T\'ecnicas (CONICET), Argentina}

\author{Alexander R. Kavner}
\affiliation{University  of  Michigan,  Department  of  Physics,  Ann  Arbor,  MI,  United  States}

\author{Ben Kilminster}
\affiliation{Universit\"at Z\"urich Physik Institut, Zurich, Switzerland}

\author{Kevin Kuk}
\affiliation{Fermi National Accelerator Laboratory, Batavia, IL, United States}

\author{H. P. Lima Jr}
\affiliation{Centro Brasileiro de Pesquisas F\'isicas, Rio de Janeiro, RJ, Brazil}

\author{Martin Makler}
\affiliation{Centro Brasileiro de Pesquisas F\'isicas, Rio de Janeiro, RJ, Brazil}

\author{Jorge Molina}
\affiliation{Facultad de Ingenier\'ia - Universidad Nacional de Asunci\'on, Asunci\'on, Paraguay}

\author{Philipe Mota}
\affiliation{Centro Brasileiro de Pesquisas F\'isicas, Rio de Janeiro, RJ, Brazil}

\author{Irina Nasteva}
\affiliation{Universidade Federal do Rio de Janeiro, Instituto de F\'isica, Rio de Janeiro, RJ, Brazil}

\author{Eduardo E. Paolini}
\affiliation{Instituto de Investigaciones en Ingenier\'ia El\'ectrica, Departamento de Ingenier\'ia El\'ectrica y Computadoras, Universidad Nacional del Sur (UNS) - CONICET, Bah\'ia Blanca, Argentina}

\author{Carlos Romero}
\affiliation{Facultad de Ingenier\'ia - Universidad Nacional de Asunci\'on, Asunci\'on, Paraguay}

\author{Y. Sarkis}
\affiliation{Universidad Nacional Aut\'onoma de M\'exico, CDMX, M\'exico}
\author{Miguel Sofo Haro}
\affiliation
{Centro At\'omico Bariloche and Instituto Balseiro, Comisi\'on Nacional de Energ\'ia At\'omica (CNEA), Consejo Nacional de Investigaciones Cient\'ificas y T\'ecnicas (CONICET), Universidad Nacional de Cuyo (UNCUYO), San Carlos de Bariloche, Argentina. }
\affiliation{Fermi National Accelerator Laboratory, Batavia, IL, United States}

\author{Iruat\~a M. S. Souza}
\affiliation{Centro Brasileiro de Pesquisas F\'isicas, Rio de Janeiro, RJ, Brazil}

\author{Javier Tiffenberg}
\affiliation{Fermi National Accelerator Laboratory, Batavia, IL, United States}

\author{Stefan Wagner}
\affiliation{Centro Brasileiro de Pesquisas F\'isicas, Rio de Janeiro, RJ, Brazil}
\affiliation{Pontif\'icia Universidade Cat\'olica do Rio de Janeiro, Rio de Janeiro, Brazil}

\collaboration{CONNIE Collaboration}

\date{\today}

\begin{abstract}
The Coherent Neutrino-Nucleus Interaction Experiment (CONNIE) uses low-noise fully depleted charge-coupled devices (CCDs) with the goal of measuring low-energy recoils from coherent elastic scattering (CE$\nu$NS) of reactor antineutrinos with silicon nuclei and testing nonstandard neutrino interactions (NSI). 
We report here the first results of the detector array deployed in 2016, 
 considering an active mass 47.6~g (8 CCDs), 
which is operating at a distance of 30 m from the core of the Angra 2 nuclear reactor, with a thermal power of 3.8 GW. A search for neutrino events is performed by comparing data collected with reactor on (2.1~kg-day) and reactor off (1.6~kg-day). 
The results show no excess in the reactor-on data, reaching the world record sensitivity down to recoil energies of about 
1~keV (0.1~keV electron-equivalent). A 95\% confidence level limit for new physics is established at an event rate of 40 times the one expected from the standard model at this energy scale. The results presented here provide a new window to low-energy neutrino physics, allowing one to explore for the first time the energies accessible through the  low threshold of CCDs. They will lead to new constrains on NSI from the CE$\nu$NS of antineutrinos from nuclear reactors.


\end{abstract}

\maketitle


\section{Introduction}
\label{sec:intro}

The Coherent Elastic Neutrino Nucleus Scattering (CE$\nu$NS) is a standard model (SM) process predicted over 40 years ago \cite{PhysRevD.9.1389}, 
shortly after the discovery of neutral-current neutrino interactions~\cite{Hasert:1973cr}. 
The coherent enhancement of the elastic scattering cross-section occurs when the energy of the scattering process is low enough and the interaction amplitude of every nucleon adds coherently to the total cross-section \cite{PhysRevD.9.1389}.
The energy at which this process dominates depends on the target nucleus. For example, for xenon the coherence requirement is $E_{\nu} < 36$~MeV, while for silicon it is $E_{\nu} < 60$~MeV.
The CE$\nu$NS has a total cross-section of $\sim 10^{-42}$ cm$^2$ for $\sim 1$~MeV neutrinos on a Si target~\cite{2015PhRvD..91g2001F}. 
Its detection was not possible until recently because of the very low energy deposition in nuclear recoils, below 15~keV for most detector targets. 

CE$\nu$NS provides a new window into the low-energy neutrino sector and the interest in this sector has been growing as a potential probe for new physics~\cite{2013RPPh...76d4201O, 2019arXiv190700991B}. 
The process is also relevant to fields beyond particle physics. For example, in astrophysics, the understanding of neutrino interactions at MeV scales is key for the energy transport in supernovae and is a limiting factor in ongoing efforts for developing new supernova models~\cite{2003PhRvDHorowitz}.
Additionally, in recent years there has been a growing interest in nuclear reactor monitoring using neutrinos~\cite{BarbeauCollar, HagmannBernstein, 2019JInst..14P6010L}. 

CE$\nu$NS from solar, atmospheric and diffuse supernova neutrinos has been identified as a limiting background for future dark matter searches \cite{2011PhRvD..84a3008A} and the 
next generation of direct dark matter detection experiments is expected to reach sensitivity to CE$\nu$NS. 
Measuring CE$\nu$NS directly in controlled neutrino experiments is needed to model and subtract this background in future dark matter experiments. 

Anomalies in reactor neutrino experiments and short baseline neutrino experiments have motivated an extension of the SM
adding a sterile neutrino~\cite{2012arXiv1204.5379A}. A number of ongoing experiments are looking to address these anomalies~\cite{2015ShortBaseline,Gariazzo:2017fdh}. CE$\nu$NS is the ideal probe to study the hypothetical sterile neutrino, because the cross-section for standard neutrinos is flavor-independent and the low energies accessible from CE$\nu$NS would allow oscillation experiments with extremely short baselines~\cite{2012PhRvD..85a3009F, Nelson:2007yq,2019arXiv190108094B,2017PhRvD..96f3013K, CANAS2018451}.

Several nonstandard neutrino interactions (NSI) predicted by extensions of the SM as well as other nonstandard neutrino properties, such as millicharge, can be probed at low energies from CE$\nu$NS~\cite{2018JHEP...07..037D,Miranda2019,2019arXiv190704942P}. For example, models in which the neutrino has an anomalous magnetic moment, whereby the neutrino-nucleus scattering is mediated by a light boson, predict a significant enhancement of the cross-section at low energies and could result in a several orders of magnitude increase in the rate of events \cite{Harnik:2012ni}. 

There are two necessary conditions for the detection of CE$\nu$NS.
The first is the availability of a source of low-energy neutrinos (below $\approx$~50 MeV) with high flux. 
The second requirement is a detector for nuclear recoils with threshold around a few keV.  
Recent technological advances in detectors for direct dark matter searches have provided several options for CE$\nu$NS detection. 
These include cryogenic bolometers \cite{superCDMS}, noble liquid detectors \cite{LZexperiment, DarkSide} and semiconducting detectors \cite{damic:2016, Abramoff:2019dfb, 2016ITNS...63.2782A}. These new detector technologies have enabled several efforts looking for CE$\nu$NS \cite{CONUS, DoubleChooz, Texono, Miner}.

Low-energy neutrinos can be produced at particle beams. 
Protons hitting a target make mesons and, if the target is large enough, the $\pi^+$ slow down and decay at rest, producing neutrinos with peak energies $\sim$20 MeV~\cite{PhysRevD.73.033005}. 
Using this technique a flux of $\sim~10^{6}$~$\nu$/cm$^2$/s/MeV is achieved at the Spallation Neutron Source (SNS) \cite{SNS}. 
The neutrinos from this source produce recoils of up to $\sim$10~keV. 
SNS produces a pulsed neutrino beam, 
which is very useful to control the backgrounds. 
The COHERENT collaboration reported the first detection of CE$\nu$NS in 2018, using a low background CsI[Na] scintillator with an active mass of 14.6~kg~\cite{Coherent}. 
These results have been used to constrain physics beyond the SM, demonstrating the potential of CE$\nu$NS as a probe for new physics~\cite{CoherentNewPhys,2018PhRvD..97c3003P,LIAO201754}.

Nuclear reactors are a powerful source of low-energy neutrinos from fission, with a flux  of $\sim10^{12}$ $\nu$/cm$^2$/s/MeV for a large reactor with thermal power of the order of $10^9$~W. 
Large reactors used for commercial power generation provide an approximately constant flux that is modulated by the nuclear fuel cycle, with typically one month shutdown every year. Smaller research reactors have a lower flux, according to their thermal power, but offer the advantage of larger flexibility in the duty cycle, providing greater control over backgrounds in the experiment. Research reactors typically allow the detectors to be located closer to their core \cite{Miner}. 

Neutrinos from nuclear reactors have an energy spectrum peaking at $\sim$1~MeV, producing recoil energies for Si nuclei with energies below 2~keV, significantly lower then neutrinos from spallation sources, making their detection more challenging. 
The search for
CE$\nu$NS in reactor experiments will allow the extension of the quest 
for new physics into the low-energy neutrino sector with sensitivity to some models that are not accessible at the energies probed at SNS~\cite{Harnik:2012ni,CANAS2018159} or providing complementary constraints to those obtained from SNS~\cite{2018PhRvD..97c5009D}.

No detection of CE$\nu$NS from reactor neutrinos has been reported yet, and the neutrino physics at this energy scale remains unexplored. Probing 
this region is the focus of the CONNIE experiment described here.

\section{The CONNIE Detector}
\label{DetectorDescription}

The CONNIE detector is an array of 14 charge-coupled devices (CCDs) operating at the Angra 2 reactor of the Almirante Alvaro Alberto Nuclear Power Plant, in the state of Rio de Janeiro, Brazil. 
The engineering prototype of the experiment was installed at the reactor site in late 2014 and the results of this run are discussed in \cite{connie:2016}. A complete upgrade of the sensors was performed 
in mid 2016, with the main objective of increasing its active mass by a factor of $\sim 40$. 
The slow control systems for the detectors were also upgraded to increase the efficiency for collecting data of scientific quality.

\subsection{CCD sensors}
\label{ccdsensors}

The CCD sensors used by CONNIE were developed by the experiment in collaboration with the LBNL Micro Systems Labs \cite{LBNLMSL}. These detectors are a spin-off from the fully depleted thick detectors that were originally designed to give astronomical instruments such as DECam \cite{2015AJ....150..150F} and DESI~\cite{2018SPIE10702E..1FM} extended sensitivity in the near-infrared region. 
CONNIE increased the CCD thickness to 675~$\mu$m. 
These are the thickest CCDs ever fabricated and are only possible to fully deplete thanks to the very high-resistivity (10 k$\Omega$-m) silicon wafers used. 
In order to reduce the thermally-generated dark current in the silicon, the sensors are cooled to temperatures below 100~K 
and operate in a vacuum ($10^{-7}$ torr).

Each sensor consists of a square array with 16 million  square pixels of 15~$\mu$m~$\times$~15~$\mu$m pitch each. 
Given their thickness, a substrate bias voltage of 70~V is applied to the backside of the detector using the method developed 
in~\cite{Holland_2003}. 
In CCDs, the charge of each pixel is usually moved towards the corners of the detectors for readout. 
In CONNIE, although the CCDs have four output stages, one in each corner of the pixel array, 
only two of the amplifiers are used in our readout setup. The charge of the full array is moved to one corner and the 16 Mpixels are read in series through a single output amplifier. At the same clock rate,
a second output of each sensor is read,  without receiving any charge from the array, to monitor the common-mode noise of the system (see section~\ref{sec:steps}).

\subsection{Packaging of the CCDs and electronics}
\label{ccdpackaging}

\begin{figure}[htb]
\centering
\includegraphics[scale=.145]{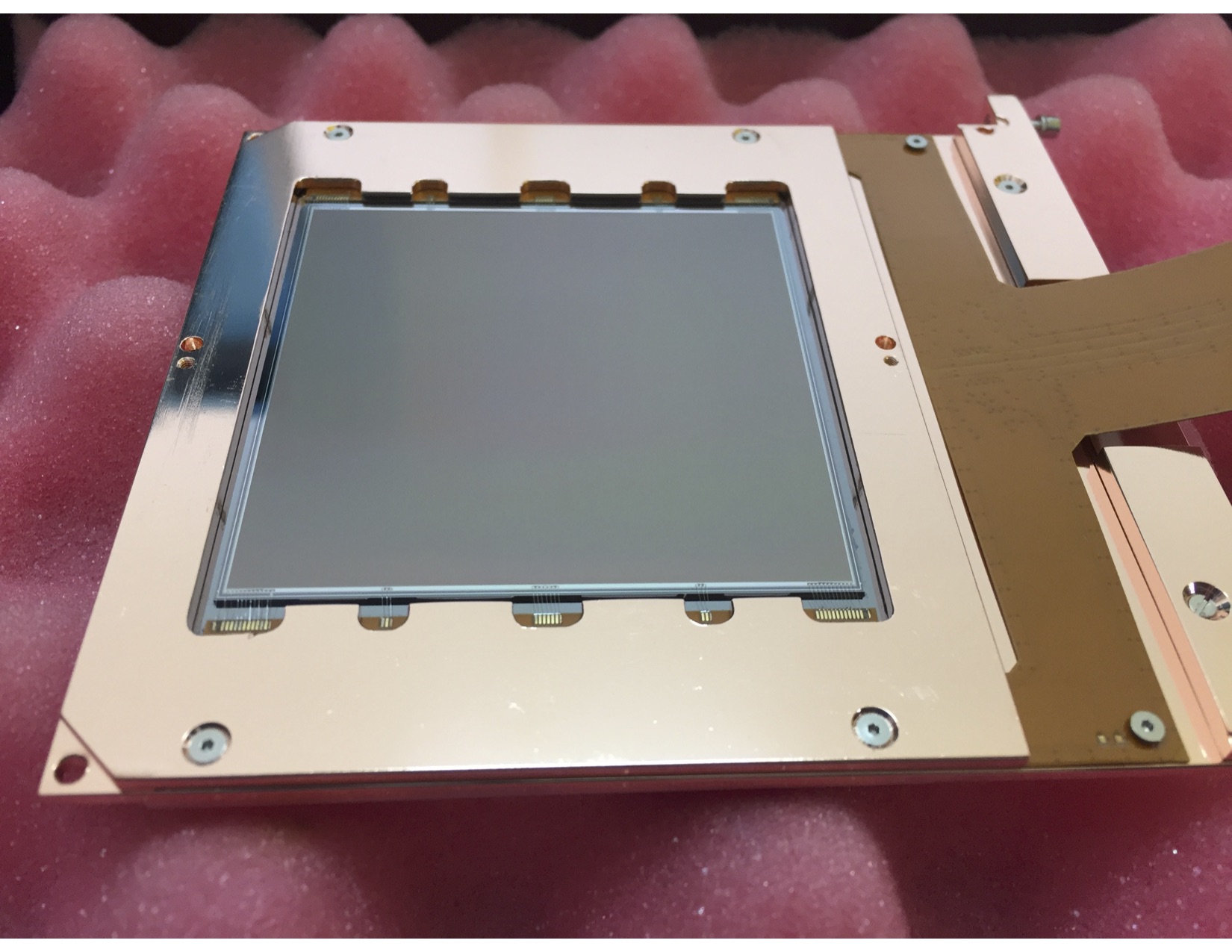} 
\caption{Image of a package including the CCD, which is glued to a silicon frame, the upper and lower copper frames and the Stage-1 flex cable.}
\label{fig:ccdpackage}
\end{figure}

Packaging of the sensors for operating in cryogenic conditions and without introducing additional sources of background is essential for low-energy measurements.
The sensors are packaged as shown in Fig.~\ref{fig:ccdpackage}. 
The back of the 6~cm~$\times$~6~cm sensor is epoxied to a slightly oversized silicon frame (7~cm~$\times$~7~cm). 
In order to avoid introducing any new materials close to the detectors, this frame is made from the same single crystal ingot 
used for the fabrication of the CCDs. 
The frame leaves most of the back of the detector exposed, only covering a few hundred rows/columns on each side. 
A flexible circuit (Stage-1) is attached to the silicon frame and micro-wire bonds are used to connect it to the pads providing the control clocks, bias voltages, and signal output of the CCD. 
The CCD sensor, frame, and Stage-1 flexible circuit are then mounted on a two-piece copper tray covering both sides of the frame, but leaving the CCD exposed.
The copper tray provides the mechanical support for the CCD package and is also the thermal connection for cooling the CCDs. 
Oxygen-free copper is used for its purity and low isotopic contamination. 

The Stage-1 circuit has no active components and serves to provide a high-density connector to the Stage-2 flexible circuit. 
The Stage-2 flexible circuit was designed for DECam~\cite{2015AJ....150..150F} and provides a source follower and preamplifiers (with gain 1.5) for the signal output. 
The Stage-2 circuit is connected to a vacuum interface board which brings the signals of all CCD packages to a 
Monsoon acquisition system \cite{2008SPIE.7014E..7OM}. The signal path after the Stage-1 circuit is exactly the same as that for the DECam imager~\cite{2012SPIE.8453E..2QS}.

\subsection{Cryogenic system} 

\begin{figure}[htb]
\centering
\includegraphics[scale=.22]{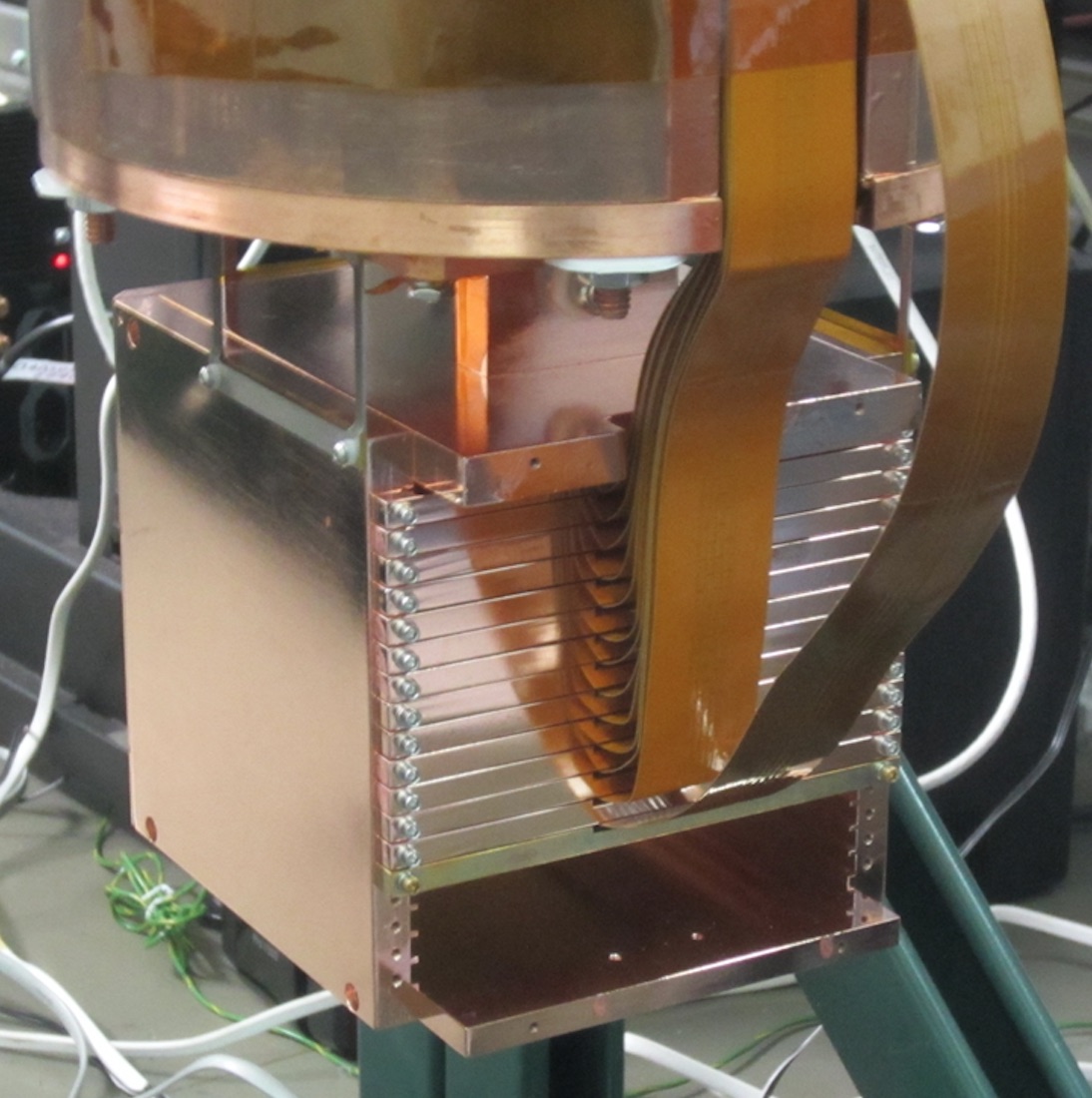}
\caption{The cold box with the 14 CCD packages installed. On top is the inner lead shield. }
\label{fig:coldbox}
\end{figure}

The array of CCD packages is mounted inside a copper cold box with capacity to hold 20 packages. Currently 14 packages are installed in the cold box (Fig.~\ref{fig:coldbox}), of which 12 are operating. The cold box is designed with the goal of shielding the sensors from any infrared radiation from the environment. It is connected to a closed-cycle helium cryocooler and the temperature of the box is controlled with a three-term controller with a precision better than  0.1 K. The cold box, Stage-1 and Stage-2 circuits are kept inside a copper vacuum vessel that is continuously evacuated using a turbo-molecular pump. The vacuum vessel is shown in Fig.~\ref{fig:shield} inside a partially-assembled radiation shield. 

\subsection{Shielding and laboratory}

\begin{figure}[htb]
\centering
\includegraphics[scale=.24]{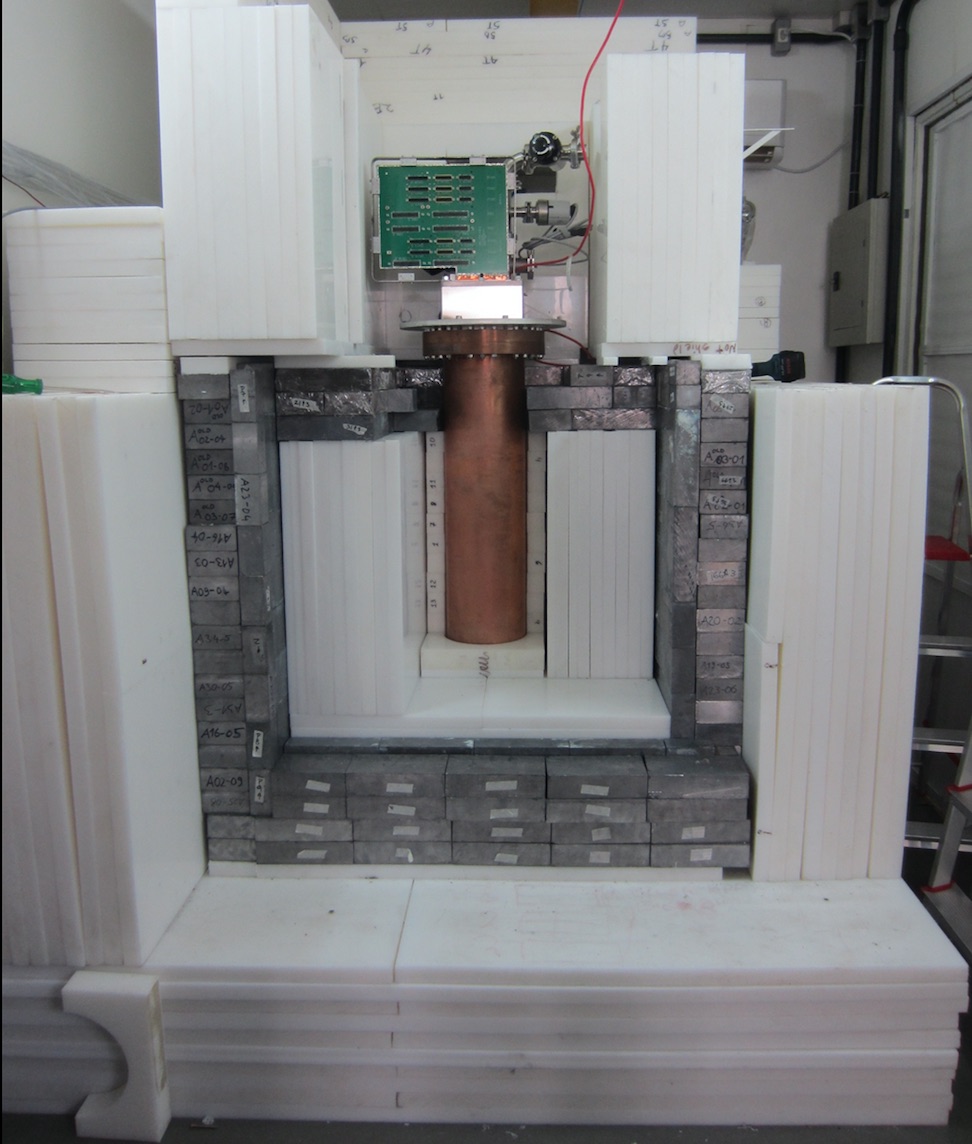} 
\caption{Image of the CONNIE detector showing the shielding partially disassembled. At the center we see the cylindrical dewar holding the copper box with the CCDs, cables, and inner lead shield. On top of it are the readout electronics. The inner and outer polyethylene layers and the lead layer of the shielding surround the detector.}
\label{fig:shield}
\end{figure}

The radiation shield is the same as in the CONNIE engineering run~\cite{connie:2016}. 
It  consists of an inner layer of 30~cm of polyethylene, followed by 15 cm of lead, and an additional outer layer of 30~cm of polyethylene  (Fig.~\ref{fig:shield}). 
Lead is a good shield for gamma radiation, while polyethylene is an efficient shield for neutrons. 
Since neutrons are produced when cosmic muons interact with lead, a fraction of the polyethylene shield is kept inside the lead layer. 
There is also a lead cylinder of 15 cm height inside the vacuum vessel, above the cold box containing the detectors (Fig.~\ref{fig:coldbox}).  
This cylinder shields the detectors from any radiation generated in the active components of the Stage-2 circuit and the vacuum interface board. 

As in the engineering run, the detector is installed inside a 
shipping container, 
located 30 meters away from the core of the  Angra 2 nuclear reactor.
The same container hosts a water-based neutrino detector,  the Neutrinos Angra experiment \cite{2019JInst..14P6010L}. 
Angra 2 is a pressurized water reactor with a thermal power of 3.8~GW
that started commercial operation during the year 2000. 
In steady-state operation, the total neutrino flux produced by the reactor is 1.21~$\times 10^{20}\, \overline{\nu}/$s \cite{2015PhRvD..91g2001F}, and the flux density at the detector 
is 7.8~$\times 10^{12}\, \overline{\nu} /\rm{cm}^2/\rm{s}$.

\subsection{Operation}
\label{Operation}

The experiment is operated remotely 
and its operating parameters and conditions are monitored and logged continuously. 
The electronic readout noise is among the most important performance parameters for the CONNIE detectors. 
This noise depends on the CCD sensors and on-chip electronics, the Monsoon readout electronics, and the interference from equipment installed inside the shipping container and outside the container. 
It is crucial to control all sources of electronic noise when the detectors are being read out. 
In order to reduce the effect of external sources of noise, all circuits, including the CCD electronics, are disconnected from the 
AC power network
when the detectors are being read out. 
This is done using an uninterrupted power supply (UPS) system that powers all the electronics connected to the detector, including the Monsoon, sensors, the computer that controls the experiment and the vacuum pump. 
During the readout stage the cryocooler is switched off, in order to eliminate the noise from its compressor.

In order to minimize the fraction of time spent reading out the CCDs and to increase the signal-to-noise ratio,  the longest possible CCD exposures are desirable. 
However, the background events (mainly cosmic muons) quickly populate the pixels. 
The exposure time was therefore chosen to keep the occupancy (fraction of pixels associated to events) below $\sim$10\%, setting that time to 3 hours. 
The duration of the readout was optimized with respect to individual pixel noise, yielding a total of 16~min 
to read the full CCD array~\cite{connie:2016}. 

\section{Image processing and catalog generation}
\label{sec:processing}

\begin{figure*}[t]
\centering
\scalebox{0.56}{\includegraphics{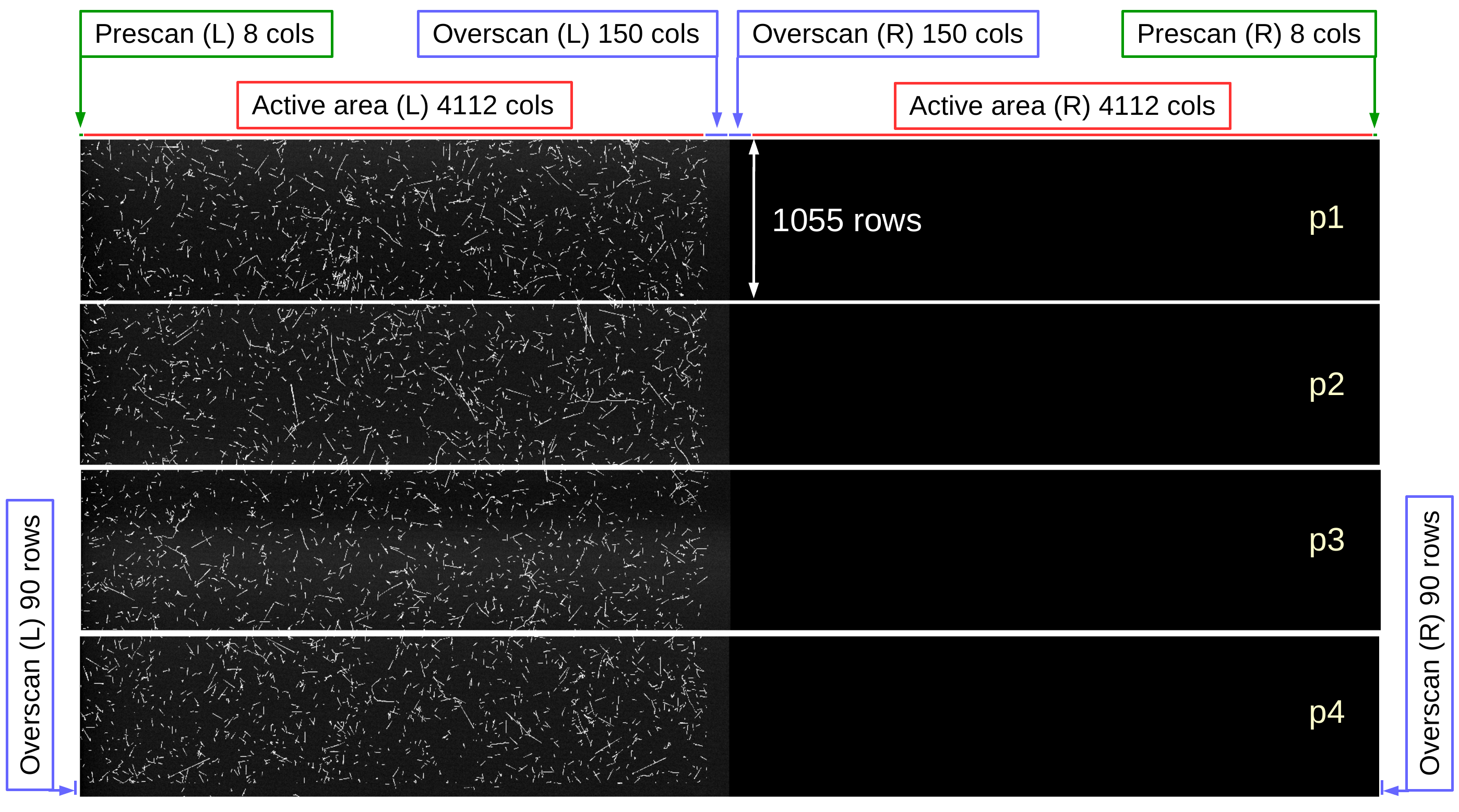}}
\caption{\small 
Schematic representation of a CONNIE image from the standard data acquisition. 
For memory handling reasons, the images are divided in 4 parts ($\tt p1-p4$) as shown here.
The charge is moved only to the left (L) amplifier, such that the right one (R) only reads the noise. A few columns with zero exposure time are read prior to moving the charge (prescan) and a larger number is read after the charge has been moved (overscan). After each column is read the readout continues for some more pixels (vertical overscan, thin strip at the bottom of $\tt p4$).}
\label{fig:split}
\end{figure*}

As mentioned in section~\ref{ccdsensors}, two output amplifiers are used 
for each CCD. The charge in the CCD is moved to the left (L) amplifier. 
The readout of empty pixels is performed on the right (R) amplifier to generate a pure noise image, at the same time as the physics data are read on L.
A few columns are read out prior to moving the charge, forming the prescan region. 
More pixel values are extracted after the charge is read, 
by overclocking the horizontal and vertical registers beyond the physical extent of the CCD, defining the overscan regions of the image~\cite{janesick2001scientific,connie:2016} (see Fig.~\ref{fig:split}).
The pixel values are recorded in Analog-Digital Units (ADU) and the data are stored as a FITS file (a standard format for CCD images \cite{2010A&A...524A..42P}). 

The data taking periods are divided into {\it runs}, which are defined as a collection of exposures that share a common detector configuration and happen during a sufficiently long and stable data-taking period. 
Some steps in the processing chain and in the energy calibration require the combination of several exposures per CCD for statistical purposes. 
A set of $\sim$60 consecutive images from the same run
provides a large enough sample for these purposes and at the same time guarantees stable conditions in the detector.
This corresponds to roughly one week of data, which provides a sufficient cadence for the data analysis and to test modifications in the data taking conditions. 
We refer to this set of images as a {\it sub-run}.
Some runs contain only one sub-run, while long and stable runs may contain many sub-runs.

\subsection{Image processing sequence}  
\label{sec:steps}

The raw images are subjected to a sequence of processing steps aimed at removing unwanted offsets and subtracting electronic noise from the pixel values. As mentioned above, the processing is carried out in batches of images that we call sub-runs.
The standard processing steps that are applied to the acquired images are: {\it i)} Overscan subtraction, {\it ii)} Master Bias subtraction, {\it iii)} Subtraction of correlated noise.

The overscan region of the CCD image is defined in Fig.~\ref{fig:split} and is used to monitor the baseline of the readout electronics. 
The overscan subtraction is an image-by-image process where the mean of the pixel values read from the overscan region is subtracted from each pixel value across the whole image. 
This has the effect of removing an image-dependent offset, making the baseline of different images comparable. 

The Master Bias (MB) is the median of all $N$ images in a given sub-run, including both the L and R sides, providing a map of the background. A set of $N$ images is formed by subtracting the MB from each of the original images, leading to ${\rm L}'$ and ${\rm R}'$.
The median absolute deviation (MAD) of the pixels from the $N$ images with respect to the MB forms the MAD image. The MAD is computed only for the physical side L and provides a measure of the width of the distribution of the values of each pixel over the $N$ images. Pixels with artificially high counts (known as {\it hot pixels}) have large fluctuations due to Poisson noise and will have large MAD values. If the MAD is much higher than the electron noise, then the pixel is flagged as misbehaving. For each sub-run we construct
a mask image with zero values corresponding to the hot pixels and one otherwise. In the data analysis, so that we consider the same pixels for all runs, a mask combining the individual masks of all runs is applied to the data.

%

Electromagnetic interference can induce noise that is correlated among the CCDs. As all sensors are read at the same time, each pixel position in the images from all CCDs in the array (and the same exposure) encodes a similar contribution of the noise from the external sources. The subtraction of correlated noise aims at removing this contribution and consists of constructing, for CCD $i$, in an array with $m$ sensors ($i=1,\dots, m$), a corrected image whose left side 
${\rm L}''_i$ is equal to the 
MB subtracted image L$_i'$  minus a linear combination of R$_k'$, i.e., 
\begin{equation}
{\rm L}''_i= {\rm L}'_i-\sum_{k=1}^{m}a_{ik}{\rm R'}_k \,.
\end{equation} 
The coefficients $a_{ik}$ are obtained from the solution of a linear system of $m$ equations with $m$ unknowns that results from requiring that the variance of the ${\rm L}'_i$ image over all the pixels is minimum: $\partial {\rm Var}({\rm L}'_i)/\partial a_{im}=0$.

\subsection{Event extraction} 
\label{cataloggen}

\begin{figure}[htb]
\centering
\scalebox{0.42}{\includegraphics{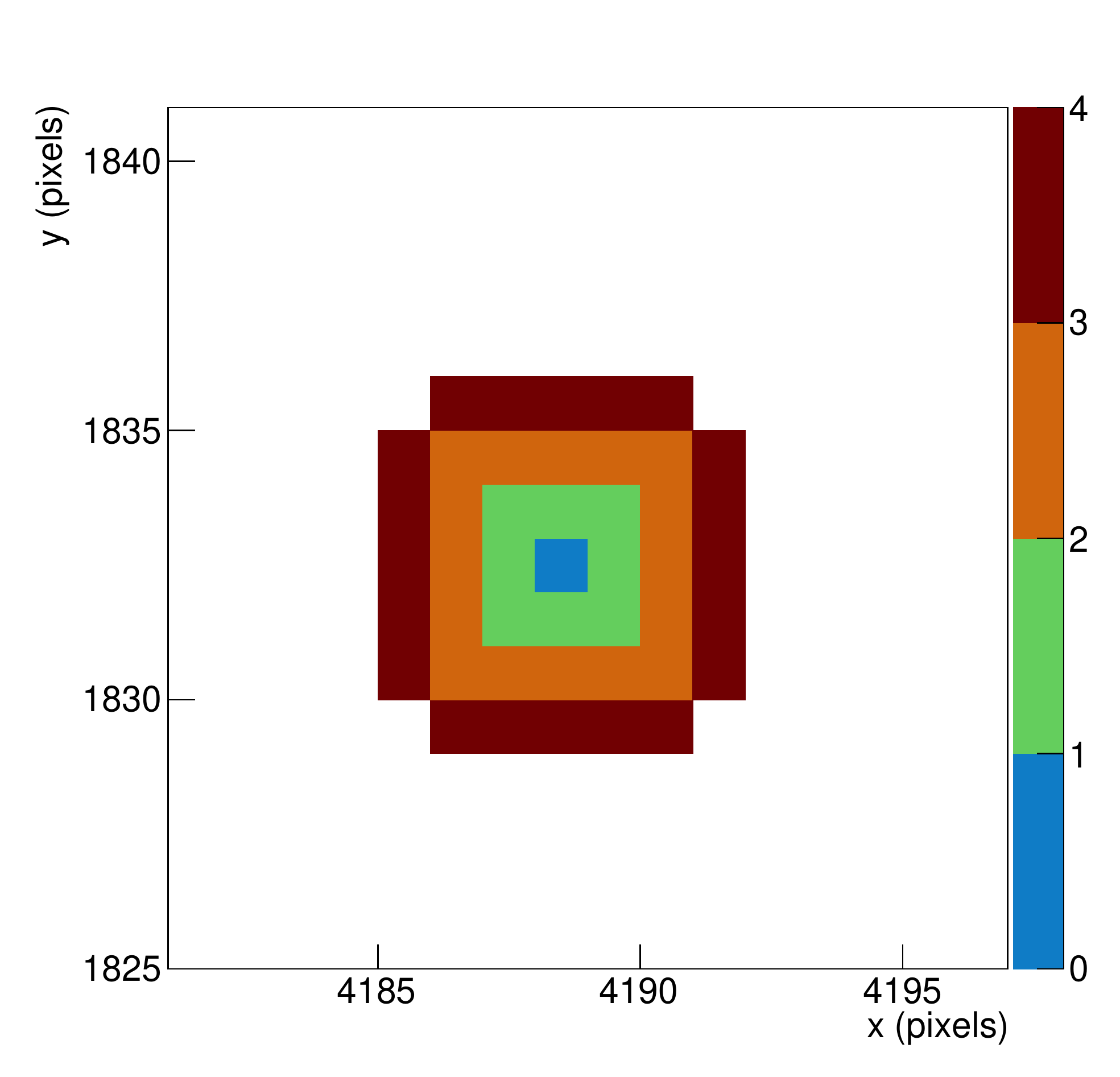}}
\scalebox{0.42}{\includegraphics{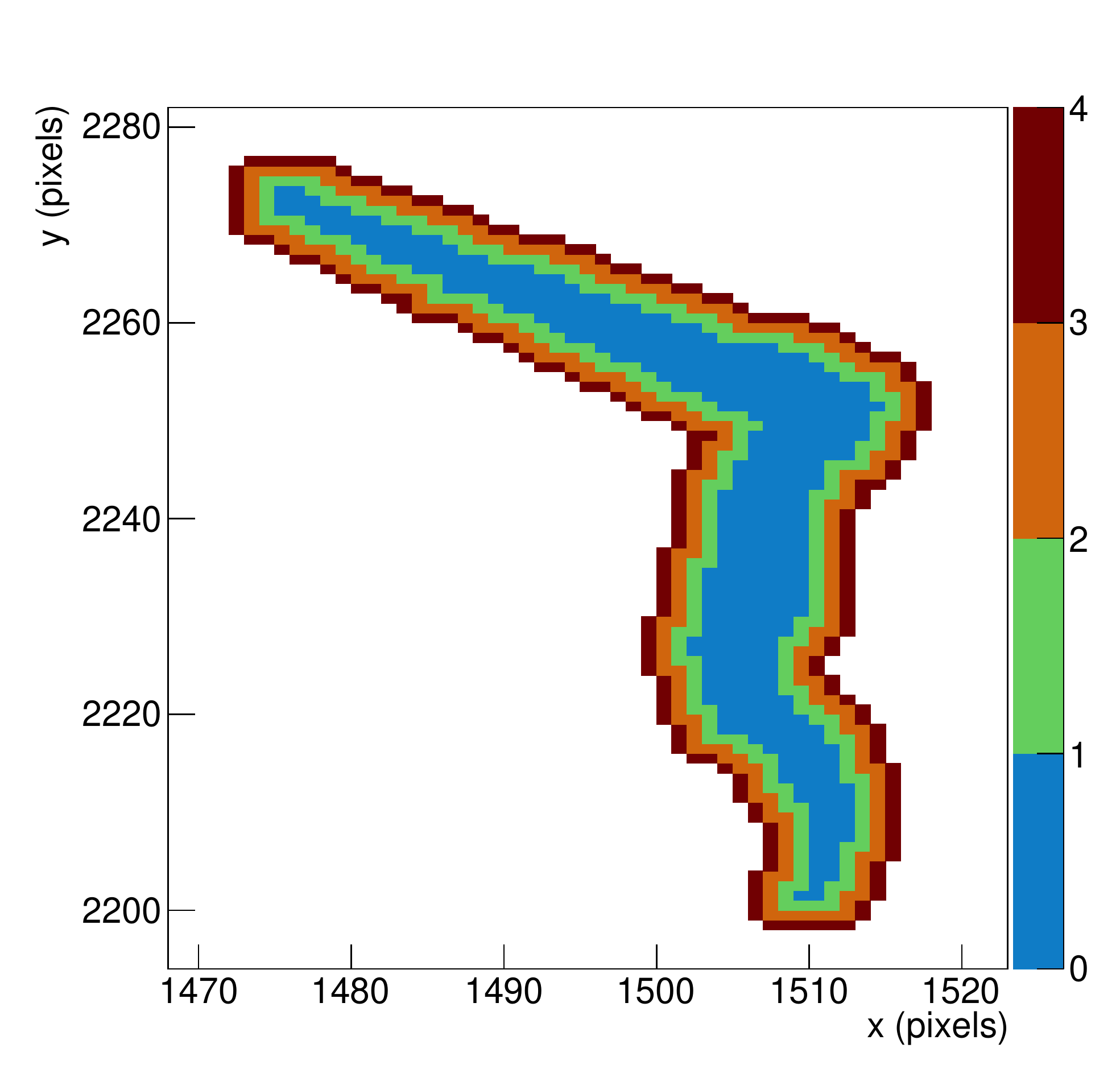}}
\caption{A cluster with one seed pixel (top) and one with 329 seed pixels (bottom),
corresponding to a muon followed by a delta ray.
The level of the pixels is given by the color scale on the right.}
\label{fig:pixel-levels}
\end{figure}

Once the final images are obtained, the next step is 
the extraction of catalogs of events, i.e., pixel clusters that are associated with energy depositions in the CCDs. 
A cluster is formed by finding its ``seed'' or ``Level 0'' pixels: adjacent pixels whose value is above a given threshold $Q_{\rm th}$ (set to 4 times a representative value for the noise in the CCDs). 
Layers of 
neighbouring pixels are then added to the seed pixels without any threshold requirement: ``Level 1'' pixels are all the pixels in 
contact with the ``Level 0" pixels, including those in the diagonal, ``Level 2'' pixels are all the pixels in contact with the ``Level 1'' pixels, and so on. 
However, for pixel levels greater than 2, only the adjacent pixels are considered. 
A cluster is the union of all pixels in all the defined layers. 
Fig. \ref{fig:pixel-levels} shows two examples of clusters with five pixel layers. 
The top panel shows an event with a single Level 0 pixel. The levels 1 and 2 include neighbouring pixels in the diagonal (forming a square shape), whereas Level 3 and above leave out the corners.
In the current CONNIE processing, the number of pixel layers is fixed to three. A catalog file containing the information of every reconstructed event in all the images in one sub-run is stored. 

As a cross check to the CONNIE event extraction pipeline, we 
have used the Source Extractor (SExtractor \cite{1996A&AS..117..393B}) code \footnote{\texttt{https://www.astromatic.net/software/sextractor}}, which is widely employed in astronomical applications. 
By adapting the SExtractor configurations to the CONNIE processed images, we obtained similar detections. 
In particular, the spectra obtained from the CONNIE extractions and the one from SExtractor
are compatible for a broad energy range. To assess the performance of the two event extraction methods at low energies we have used images containing simulated neutrinos, which are produced to determine the reconstruction efficiency (see Sec. \ref{efficiencySec}). We found that SExtractor generates spurious detections for energies $\lesssim$0.2~keV, which includes the energy range in which we are interested in this paper, whereas the CONNIE extraction does not have such a problem. Therefore we consider only the CONNIE extraction in the rest of the analysis.

\section{Calibration of the sensors} 

\subsection{Energy calibration} 
\label{calibsec}

As described in section~\ref{ccdpackaging}, the CCDs are attached to a frame 
made of the same high-purity silicon as the detectors themselves and are embedded in a copper-rich environment. 
Therefore, the emission of Cu and Si fluorescence x-rays from excitations by cosmogenic particles and gammas from inherent radioactivity is readily observed in all the sensors as peaks in the energy spectrum. 
The two principal Cu fluorescence x-rays have energies of 8.047 keV (K$\alpha$) and 8.905 keV (K$\beta$), while the Si fluorescence x-rays have an energy of 1.740 keV. 
These peaks provide a way to monitor the detector calibration continuously.
The linearity and resolution in the energy response of these CCDs have been thoroughly characterized down to energies of 
$\sim40~{\rm eV}$ in previous work \cite{damic:2016,Chavarria:2016xsi}. 

\begin{figure}[thb]
\includegraphics[scale=0.39]{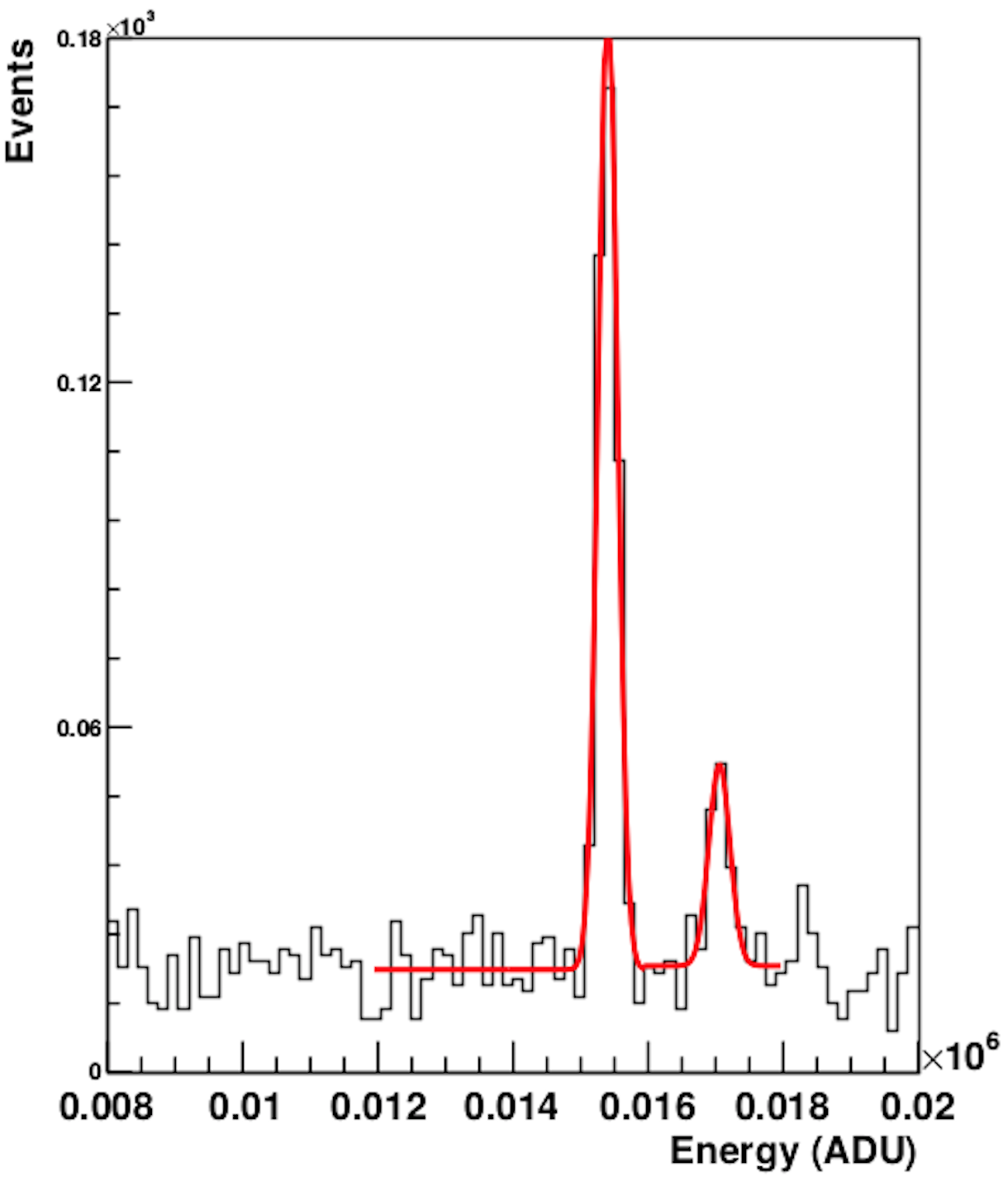} 
\caption{Energy spectrum (in ADU) around the copper fluorescence peaks for 60 consecutive three-hour exposures. 
The first peak, corresponding to K$\alpha$ at 8.047~keV, is fitted by a 
Gaussian and its mean is used to obtain the calibration constant. 
The second peak is fitted by the same function.}
\label{fig:0Y}
\end{figure}

\begin{figure}[htb]
\includegraphics[scale=0.41]{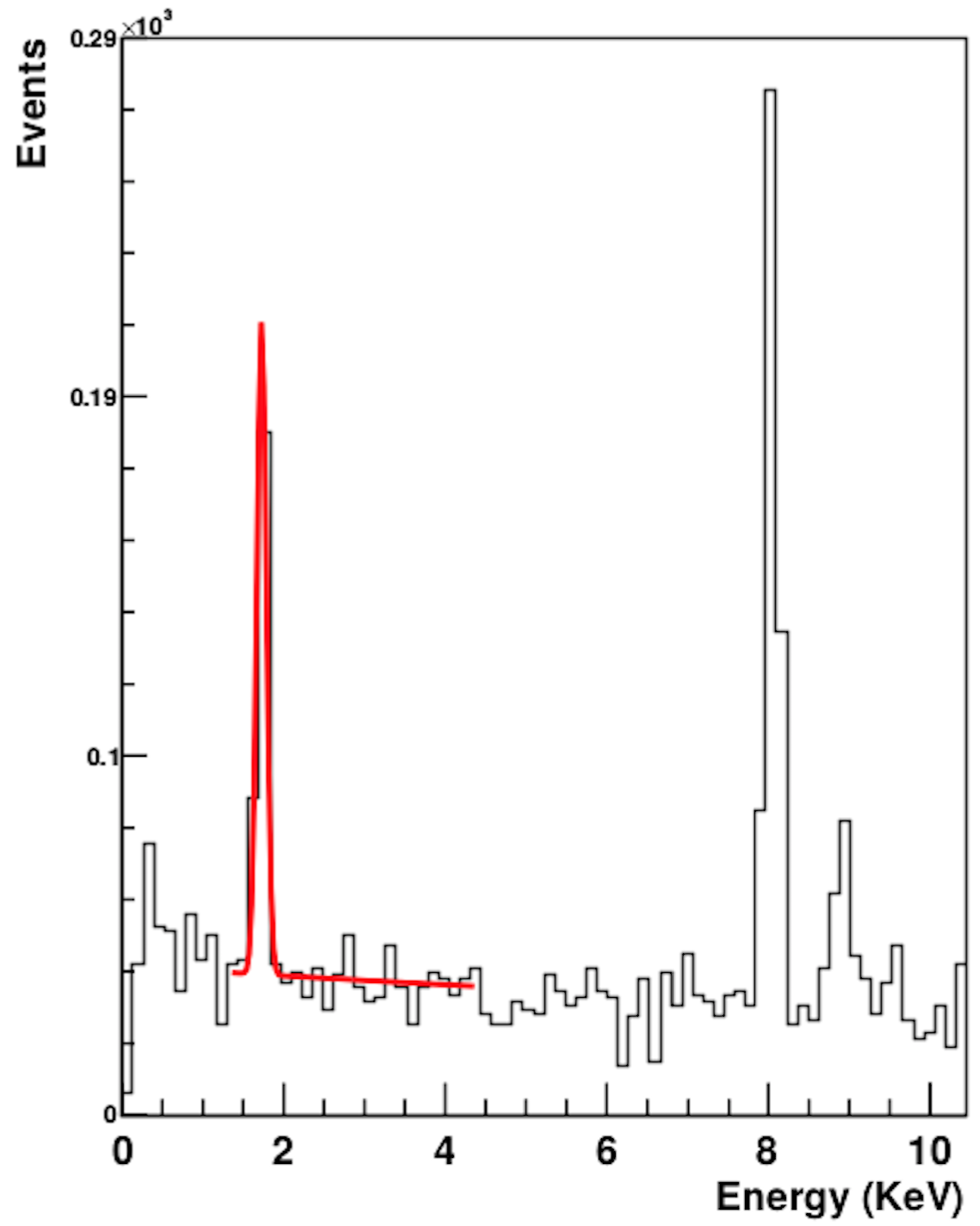} 
\caption{The calibrated energy spectrum in the region up to 10.5~keV. 
The Si fluorescence peak at 1.740~keV is fitted by a Gaussian plus an exponential.}
\label{fig:0Y2}
\end{figure}

The energy in ADU is defined as the integrated value of all pixels in a cluster on the processed image down to a given level. A calibration constant (in keV/ADU) is calculated for each CCD in every sub-run of $\sim$60 images using the Cu K$\alpha$ peak.  
Fig. \ref{fig:0Y} shows the region of the spectrum around the two Cu fluorescence lines for all the events in a sub-run. 
Fig. \ref{fig:0Y2} shows the calibrated spectrum for the same sub-run in the energy range (0--10.5) keV where the Si fluorescence line is visible.


The stability of the calibration within sub-runs was monitored by looking at the position of the Cu-K$\alpha$ peak in groups of five consecutive images, fitted with a Gaussian plus constant-background model. 
The calibration constants extracted from these smaller groups of images were found to be stable within 0.2\% over periods of time extending for several months.

\subsection{Energy resolution} 
\label{sec:eresolution}

The energy resolution for photons is a well-understood quantity for CCD sensors: at high energies (several keV) the energy resolution is dominated by the silicon ionization efficiency which is proportional to the energy of the photon through the Fano factor \cite{janesick2001scientific}. 
This factor was evaluated in the laboratory, for the same type of CCD as CONNIE, using x-rays (typically of 5.9 keV from a $^{55}$Fe source) 
giving a value of 0.133 \cite{damic:2016}. 
At low energies (below 0.1~keV) the energy resolution is dominated by the readout noise of the sensor and is evaluated by adding low-energy events to the data 
and measuring the energy dispersion for those events after reconstruction, which is found to be 0.034~keV. 
The total energy resolution for photons is the sum of both effects and can be modelled by a normal distribution with variance
\begin{equation}\label{eq:energyResolution}
    \sigma_{\rm{res}}^2=\left(34 \,{\rm eV}\right)^2 + \left(3.745 \, {\rm eV}\right)FE\,,
\end{equation}
where 3.745~eV 
is the adopted mean ionization energy required to produce an electron-hole pair for photons taken 
from \cite{1973ITNS...20..473R}, $F$ is the Fano factor and $E$ is the photon energy in eV. 

\subsection{Size-depth calibration}
\label{sec: size calibration}

The event shape in the data depends on the transport of charge carriers in the depleted silicon before they are trapped by the potential well of each pixel. 
Once the free carriers are generated, they drift under the electric field of the depleted silicon. 
This electric field has only one component transverse to the array plane and free holes have essentially no restriction to move laterally before being trapped by the pixel well. 
The magnitude of the lateral dispersion is determined by the drift time, which is set by the distance from the primary ionization point (depth in the silicon) to the well positions (approximately 2~$\mu$m below from the front face of the sensor). 
In particular, neutrinos deposit such a small amount of energy when scattering off nuclei that the primary ionization volume is much smaller than the subsequent dispersion of the free carriers. Measuring this process is needed for the complete characterization of 
the shape of the neutrino events and is a key ingredient for the simulations used to characterize the reconstruction strategy. 

The lateral dispersion produced by thermal diffusion follows a Brownian motion with a position probability that can be modeled by a two-dimensional Gaussian distribution \cite{Holland_2003} with equal standard deviation ($\sigma_D$) in both directions, given by
\begin{equation}
    \sigma_D^2 = \alpha \ln(1-\beta\, z),
    \label{eq:event size}
\end{equation}
where $\alpha$ and $\beta$  are parameters that condense several physical constants of the sensor and $z$ is the depth of the interaction in the bulk of the silicon. 

Since the parameters in Eq.~(\ref{eq:event size}) depend on fabrication and operation parameters such as the doping content, sensor thickness and substrate voltage, we measure them independently for each detector from the data using a high-purity sample of atmospheric muons.
They produce a continuous ionization trace in the output data with a very high probability of crossing the entire detector thickness. 
Furthermore they leave a nearly constant energy deposition per unit length, allowing us to assume a uniform initial ionization cloud radius along the track length.

\begin{figure}[htb]
\centering
\scalebox{0.4}{\includegraphics{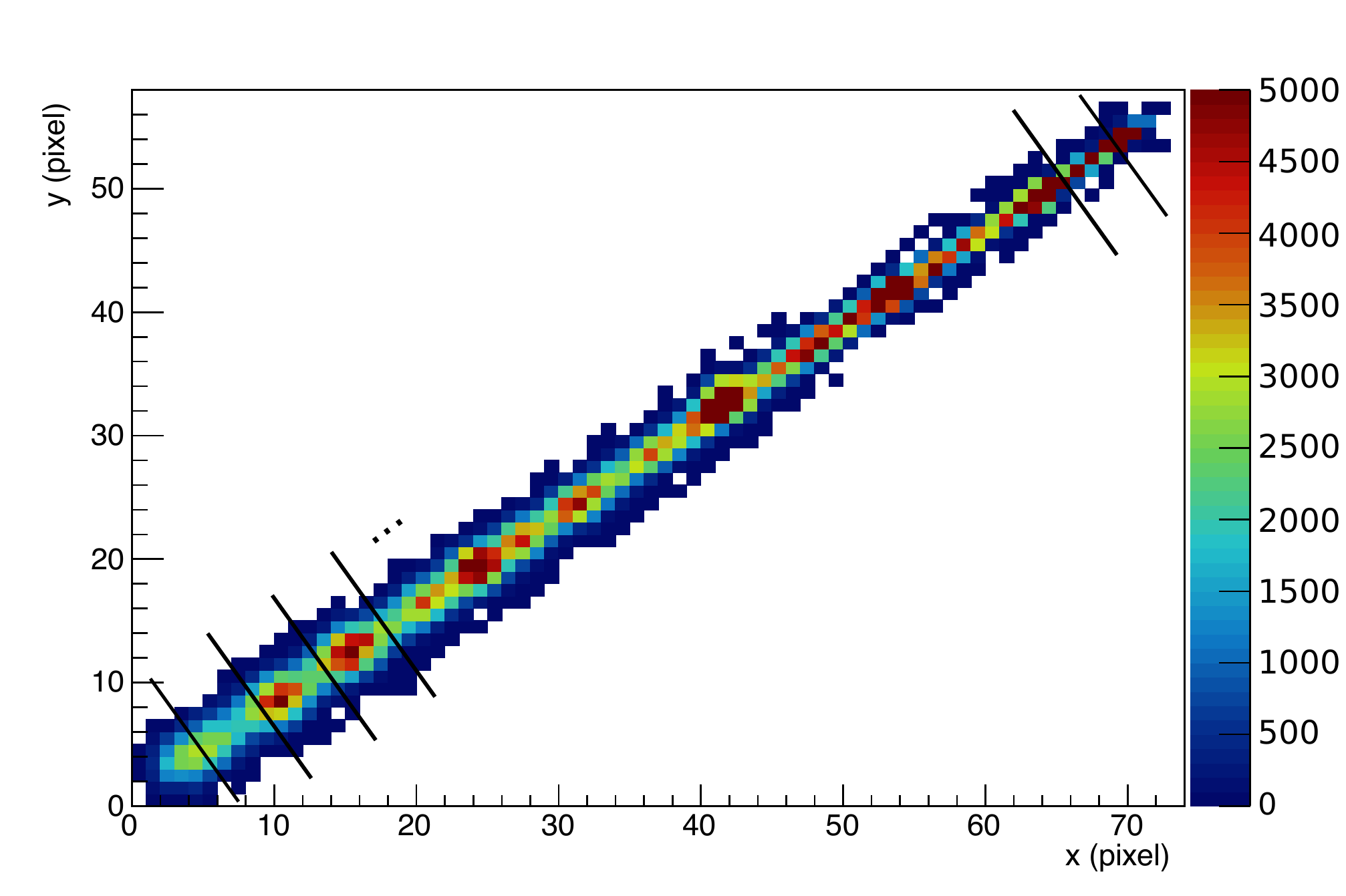}}
\caption{A typical muon event crossing the entire thickness of the detector where the color scale indicates the charge value of each pixel in arbitrary units. The lines perpendicular to the track define the segments where the transverse standard deviation is measured. }
\label{fig:calibration with muons1}
\end{figure}

The muons are selected as straight tracks with a 2D projected length over 150 pixels and for which both ends are not at the edges of the CCD active area, such that they traverse the full thickness of the CCD. This selection was tested using visual inspection and yields a less than 1\% contamination.
An example of a muon track is shown in Fig.~\ref{fig:calibration with muons1}.
These straight events 
have a thicker end,
corresponding to ionization produced in the back of the sensor and a thinner end from ionization at the front (close to the pixel potential well). 

\begin{figure}[htb]
\centering
\scalebox{0.43}{\includegraphics{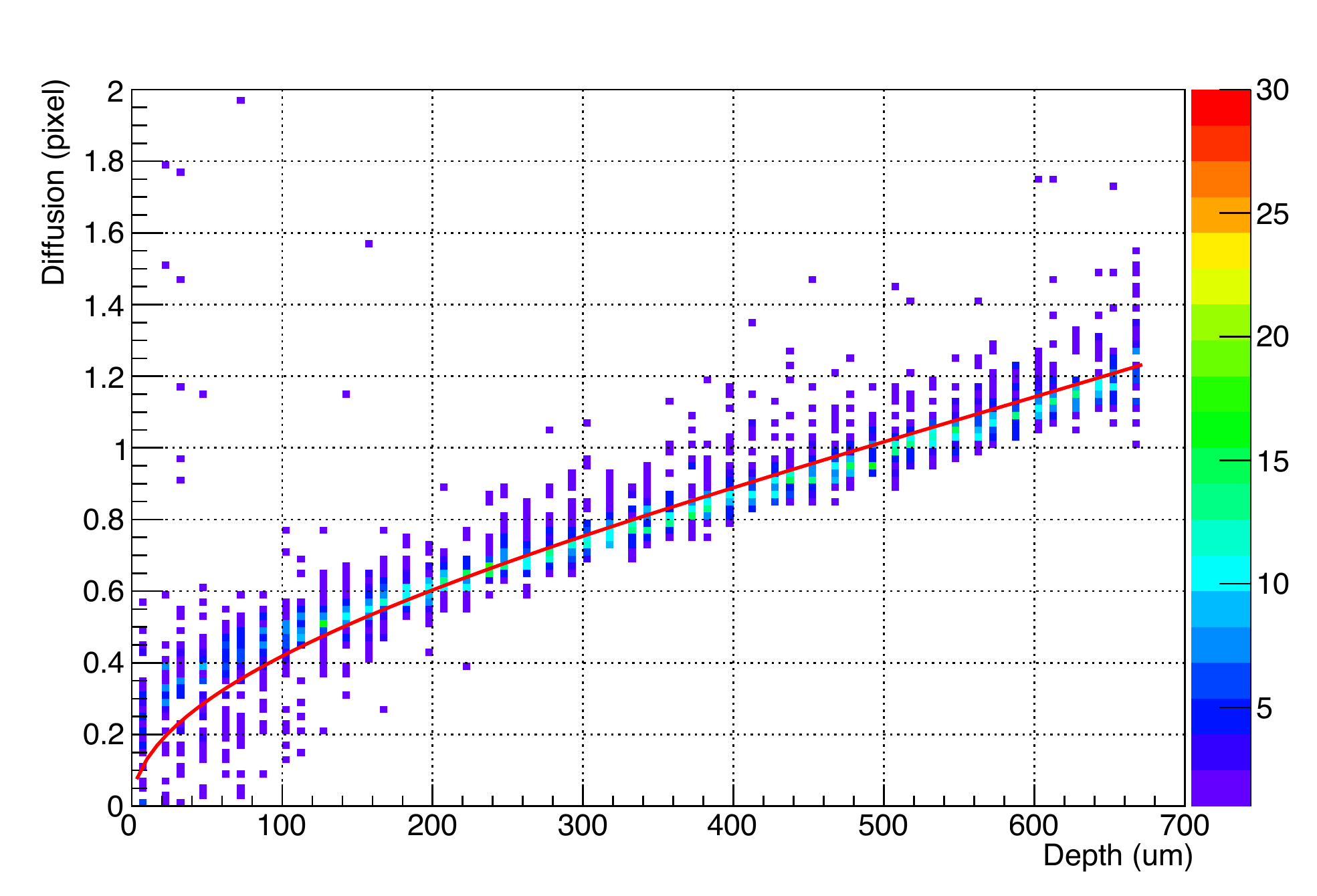}}
\caption{Two-dimensional histogram illustrating the measurement of the size-depth relationship in one of the sensors.
The vertical axis shows the width of the muon segments for each depth in the horizontal axis as measured from muons traversing the CCD.
The color scale indicates the number of measurements that lie in each diffusion-depth bin.
The solid line is the best fit of Eq.~(\ref{eq:event size}) to the distributions.}
\label{fig:calibration with muons}
\end{figure}

The width of the track can be mapped as a function of the distance from the first interaction point, 
which is proportional to the depth of the hit 
in the CCD. 
In practice, the track is divided into segments where the transverse standard deviation is measured. 
An example of this mapping using many muons is shown in Fig.~\ref{fig:calibration with muons}. 
Due to the kinetic energy of the liberated electrons, there is a nonzero size for the initial ionization cloud, which in our data is of the order of 0.1 pixel. We perform a fit to Eq.~(\ref{eq:event size}) for depths above $\simeq~100~\mu m$ to minimize the pixelation effects.
%
The average values for the detector array are
$\alpha = -690~\mu \rm{m}^2$ and $\beta = 6 \times 10^{-4} /\mu$m, 
with a 10\% dispersion between detectors and  a maximum diffusion width of $\sigma_D=19$~$\mu$m.

\section{Detector performance}\label{Sec:detPer}

\subsection{Stability of the background radiation}

\begin{figure}[t]
\centering
\scalebox{0.46}{\includegraphics{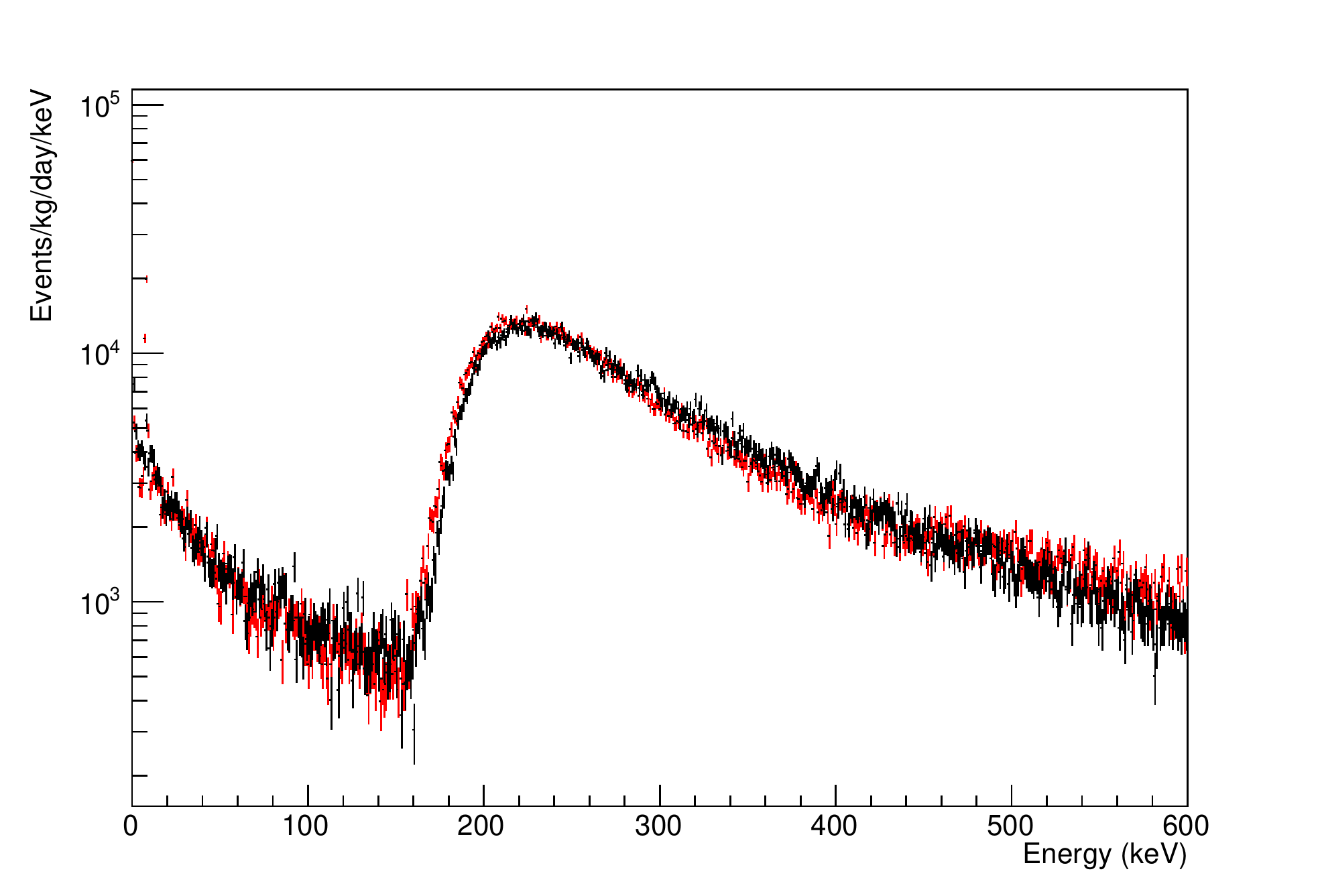}}
\caption{Comparison of the simulated spectra from muons and data in one of the sensors. The red curve represents data from one of our runs (0.048 kg-day), while the black curve represents the muon contribution obtained with the Geant4 package. The simulated spectrum is based on the expected rate of muons at sea level and has not been fitted to the data.}
\label{fig:muon_comp}
\end{figure}

The full energy spectrum for the data collected in the CONNIE experiment is shown in Fig.~\ref{fig:muon_comp} (red curve).
The excess at around 250~keV corresponds to minimum ionizing cosmic muons traversing the silicon sensors. The 
increase in the rate at lower energies is dominated by secondary products of these muons, showering in the detector and in the nearby shielding components. 
This is demonstrated with a full Geant4 \cite{AGOSTINELLI2003250}
simulation of atmospheric muons hitting the detector, following the energy and angular distribution of~\cite{smith-duller:1959}, which reproduces reasonably well the moun distribution at sea level~\cite{chatzidakis:2015}. 
The resulting spectrum matches the shape of the overall CONNIE spectrum above 10~keV, as shown in Fig.~\ref{fig:muon_comp}  (black curve) and 
gives an overall background rate consistent with the total rate observed. 
A more sophisticated background model including all the low-energy processes below 10~keV is left for future work, but we know that at least a fraction of the background in this energy range is also due to the secondary products of the muons. Therefore, to search for low-energy events it is important to monitor the stability of the muon background and its low energy by-products.

We monitor the stability of the background by looking at the rate at the fluorescence peaks as well as in regions away from the peaks. As our CE$\nu$NS analysis is based on a comparison of reactor on and off data, 
we study the stability by grouping the data collected during these two states of the reactor
(the selection of the specific periods used in this analysis is presented in Sec.~\ref{dataselection}).

\begin{figure}[htb]
\scalebox{0.53}{\includegraphics{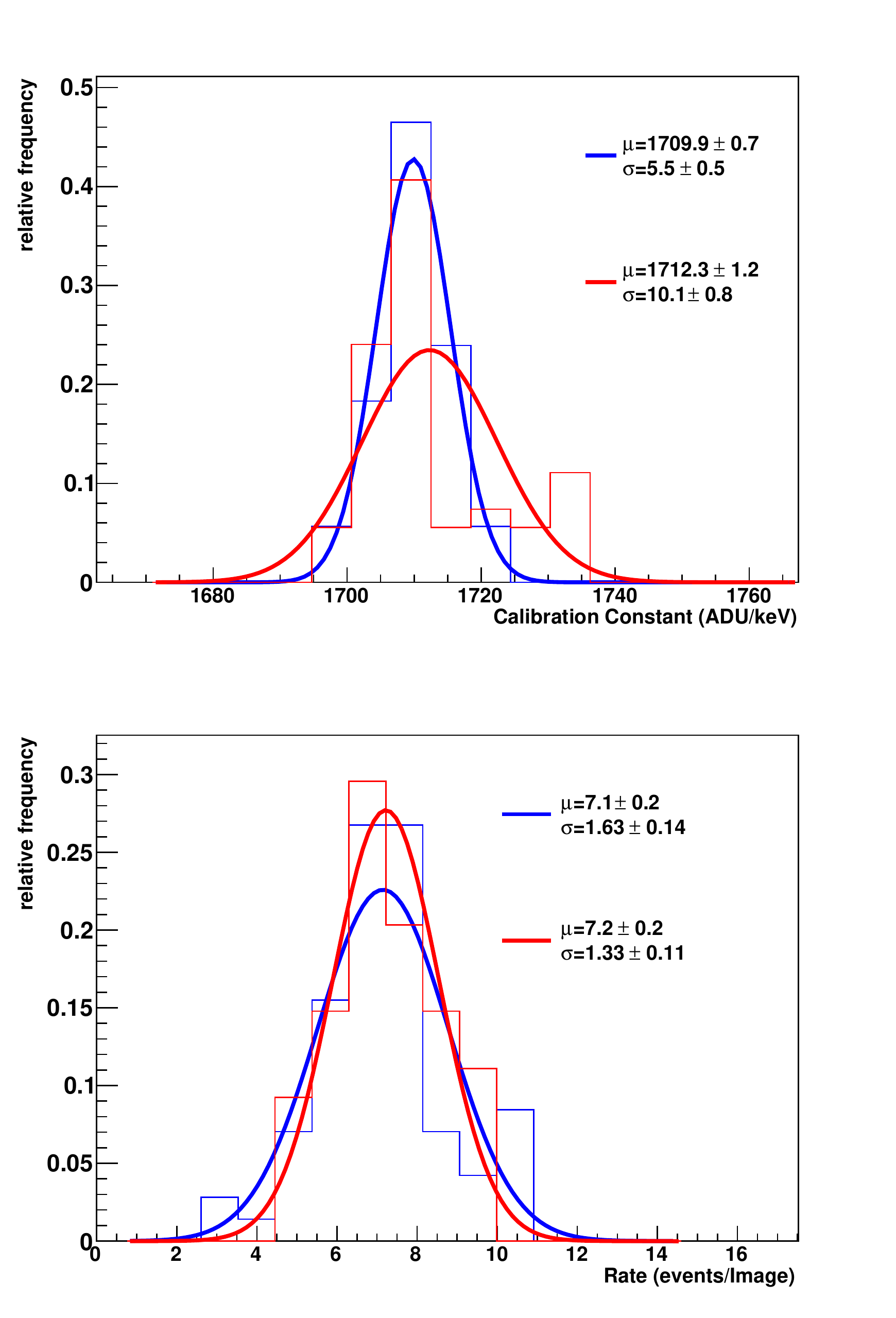}}
\caption{Top: Unit-normalized distributions of the fitted calibration constant (Cu K$\alpha$) from spectra of groups of 5-6 images, for the reactor-on (blue) and reactor-off (red) periods. Bottom: unit-normalized distributions of the event rate under the Cu K$\alpha$ peak 
calculated as the area of the fitted Gaussian. 
The mean ($\mu$) and width ($\sigma$) of the distributions are also shown.}
\label{Fig:1Y}
\end{figure}

As discussed in Sec.~\ref{calibsec} the copper peaks are fitted by a Gaussian plus a constant, providing the energy calibration (position of the peak) and also the event rate integrated on the peak. In Fig.~\ref{Fig:1Y} we show the distribution of these two quantities for the Cu K$\alpha$ peak for exposures during the periods of reactor on and off. As mentioned before, we see that the calibration is extremely stable during the operations and is independent of the reactor state. The rate is also very stable, with fluctuations consistent with Poisson statistics and no significant difference between the reactor on and off periods.

The position and rate of the Si fluorescence peak in the low-energy region of the calibrated energy spectrum were also monitored and found to be consistently stable. 
The Si peak is also stable comparing the on and off periods,
with a mean of 1.738 (1.736)~keV (using the Cu K$\alpha$ calibration) and width 
of 0.001 (0.003)~keV during the on (off) period.
%

\begin{figure}[htb]
\begin{center}
\includegraphics[width=.49\textwidth]{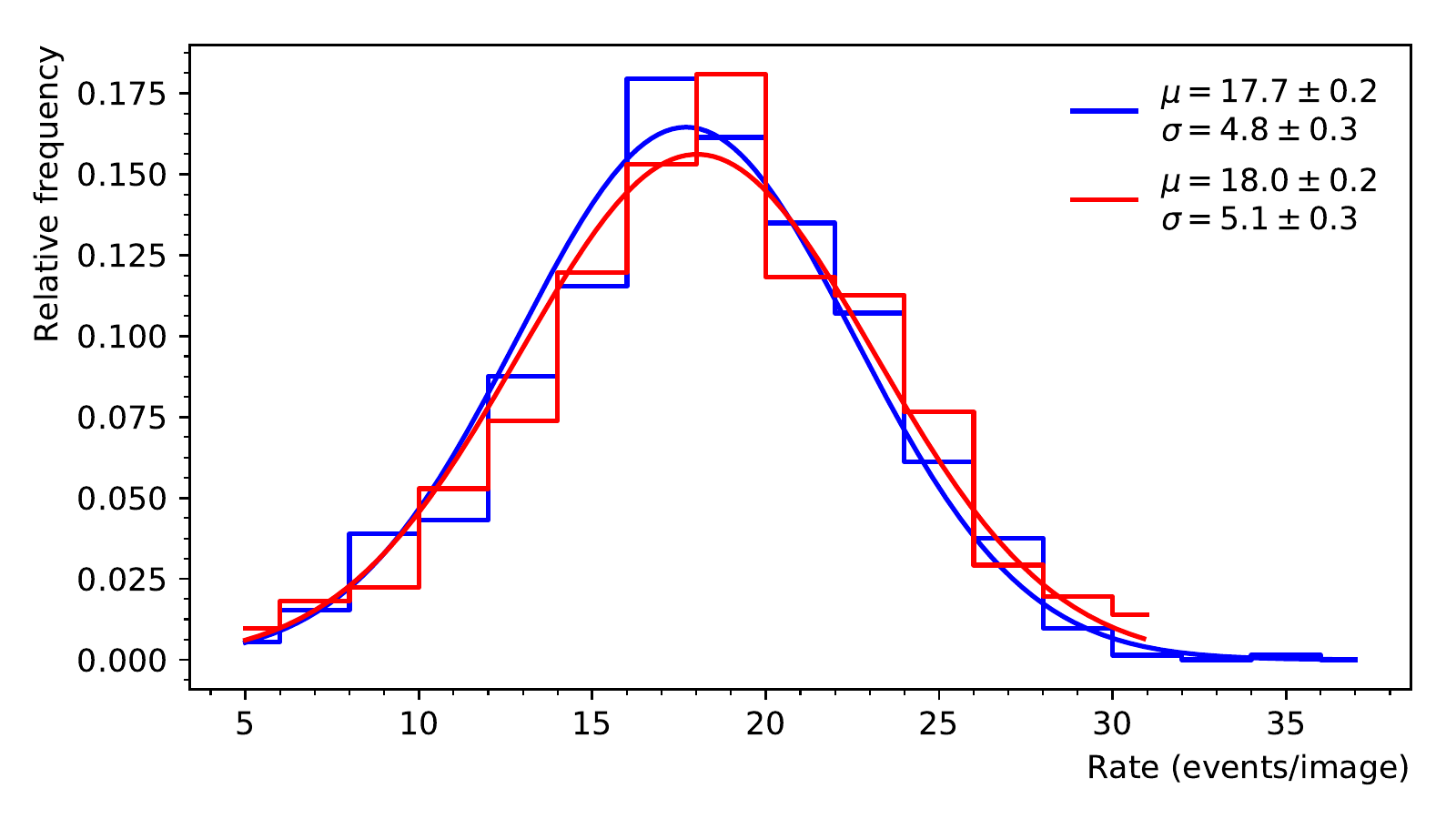}
\includegraphics[width=0.49\textwidth]{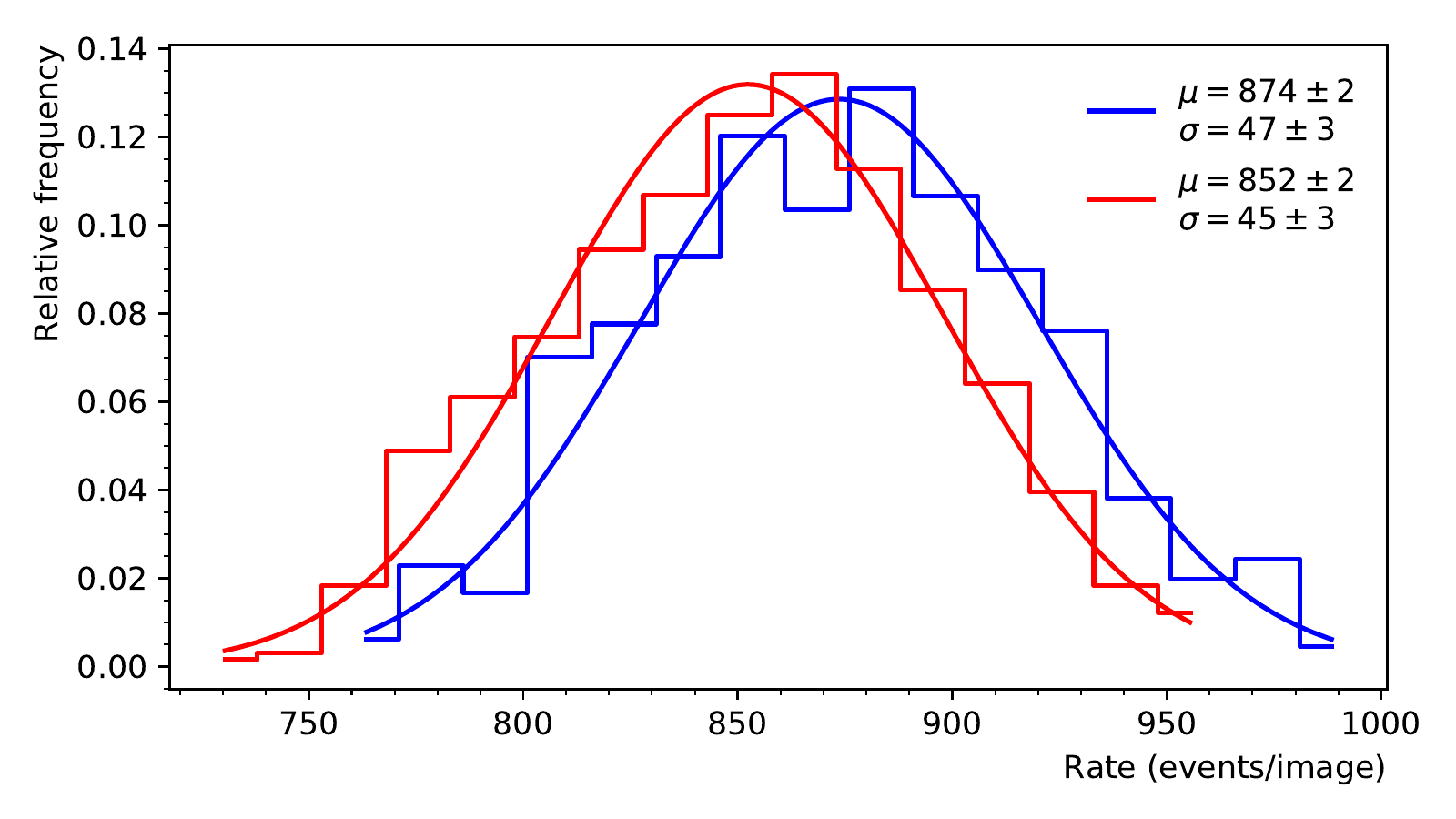}   
\caption{Unit-normalized histograms of events per image in the energy ranges 3--7~keV (top) and 250--350~keV (bottom), for the reactor on (blue) and off (red) periods. The mean ($\mu$) and width ($\sigma$) of the distributions are also shown.}\label{fig:counts}
\end{center}
\end{figure}

We have also monitored the stability of the background radiation in two energy ranges.
The first, from 3 to 7~keV, was chosen to be between the Cu and Si fluorescence peaks and is dominated by low-energy photons. The second range, from 250 to 350~keV, was chosen to include the muon peak and, thus, is dominated by muon events. Fig.~\ref{fig:counts} shows the distribution of the event rate in one of the sensors in these two energy ranges for
the periods of reactor on and off. 
For the low energy range, the distributions show that the radiation background is constant on the two periods, within statistical uncertainties.
In the high-energy region, we notice a 2.5\% variation in the rate, but it does not affect the low energy spectrum, at least at the level of precision achieved in this paper. The differences appear not only in the on versus off periods, but have a long term variation along the months.

\subsection{On-chip noise sources}  
\label{RNDC}

On-chip noise sources are the main contribution to the measurement error in the pixels \cite{janesick2001scientific}.
The dominant effects are produced by two independent processes: the readout noise (RN) added by the output amplifier of the sensor to the output signal and the dark current (DC) which is a spurious generation of charge by 
thermal excitations in the crystal. 
The RN follows a Gaussian distribution while the DC follows a Poisson distribution \cite{janesick2001scientific}. 
Both quantities are constantly monitored in the experiment by measuring the parameters of their distributions with fits to the combined probability function of the pixels without events for each image. 
Since both noise sources are independent, the joint probability function for the energy in a given pixel due to the combined DC plus RN processes ($f(E)$) 
can be calculated as the convolution of the marginal contributions:
\begin{equation}
f(E) = \sum_{n=0}^{\infty} \frac{1}{\sqrt{2 \pi}\: \sigma g}\exp\left(-\frac{(E-ng +\lambda g - \mu)^2}{2\left(\sigma g\right)^2}\right) \frac{\rm{e}^{-\lambda} \lambda^n}{n!},
\end{equation}
where $E$ is the energy in ADU, $g$ is the gain of the system in units of ADU/e$^-$ calibrated with x-ray lines, $n$ runs 
over all the possible numbers of generated charges from the DC process, $\lambda$ is the mean number of generated charges per pixel by the DC, $\sigma$ is the standard deviation of the RN process in units of e$^-$, and $\mu$ is an external parameter added to account for a small remnant in the baseline subtraction processing step. 
Note that the central value of each term in the sum is adjusted by $\lambda g$, which is not part of the original Poisson marginal probability function. 
It is included in order to correct for the effect of subtracting the median image in the processing chain. 

\begin{figure}[ht]
\includegraphics[scale=0.254]{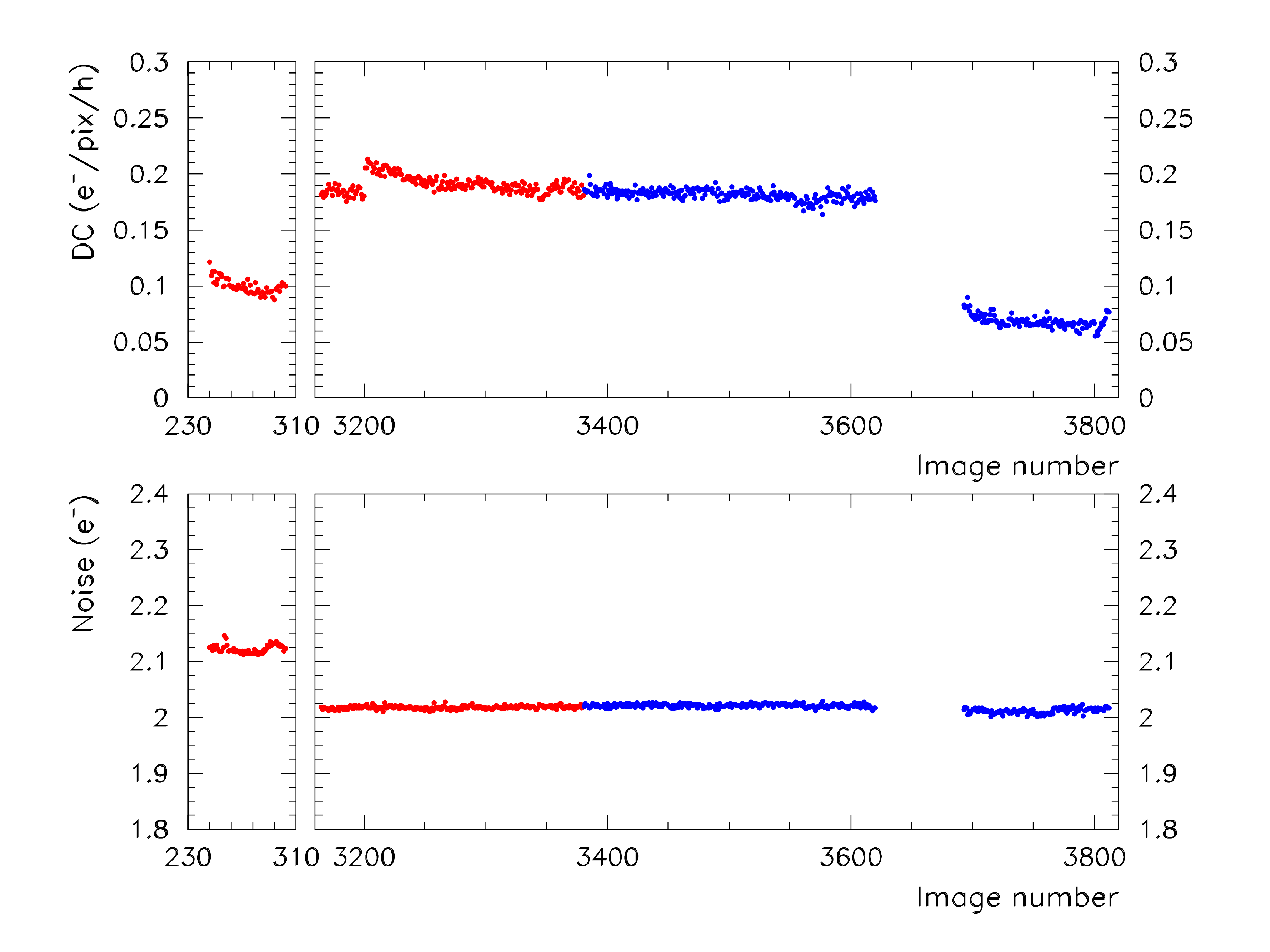}
\caption{On-chip noise components in one of the sensors for all exposures considered for this paper. Each point corresponds to a three-hour exposure. The red points are taken during reactor off periods, while the blue ones are during on periods.}
\label{fig:dcrn}
\end{figure}

The method was used to evaluate the noise sources on each output image. The distribution of DC 
($\lambda$) and RN 
($\sigma$) for a single CCD
is shown in Fig.~\ref{fig:dcrn} for all exposures considered in this paper (see Sec.~\ref{dataselection}).
For all sensors the typical ranges are:
$\lambda \simeq 0.05-0.25$~e$^-$/pix/hr and $\sigma \simeq 1.7-2.2$~e$^-$.
As we see in this figure, the DC had significant changes along the operations of the detector (see Sec.~\ref{dataselection}). The low-energy event selection 
(Sec.~\ref{eventselection}) is constructed so that these variations do not affect the reactor-on versus reactor-off comparison.

The effect of these noise sources in the event selection is different: the RN fluctuations dominate the root mean square (RMS) error of the pixel and have an impact on the energy resolution of low-energy events, while the DC, with a larger-probability tail for positive values, has more impact on the number of spurious (false positive) events. 
These two mechanisms are taken into account for the event selection cuts, as described in Sec.~\ref{eventselection}.

\section{Event selection and efficiency} 

\subsection{Data quality selection criteria}
\label{dataselection}

The detector has been taking data continuously since August 2016, with short interruptions due to planned on-site interventions (repairs and upgrades in the control system and container infrastructure) or power cuts. 
The data collection can be divided in 3 seasons. The first, from August 2016 to March 2017, includes one of the reactor shutdowns and ended with the planned period of maintenance of the detector. The second one, from March to December 2017, does not include any reactor off period and was defined by the infrastructure upgrade of the lab. The third season, from January to August 2018, includes the second reactor shutdown used in this measurement. 

The operating conditions varied between the seasons, leading to modifications in the DC and RN. For the analysis presented in this paper we consider data that were acquired in the first and third seasons of the experiment, when the CCDs used in the analysis have a good 
performance. 
More specifically, we required RN 
better than 2.2~e$^{-}$ and DC 
less than 0.3~e$^{-}{\rm /pix/h}$.

The two selected seasons correspond to the different panels in the DC and RN plots of Fig.~\ref{fig:dcrn}. A gap is seen in the third data-taking season corresponding to a 10-day period of thermal problems in the container, with data that does not pass our quality cuts. After new operating conditions and cleaning the charge in the CCDs, the DC had a substantial improvement.
The total period considered in this work corresponds to 6 sub-runs of reactor-on data and 5 sub-runs of reactor off.


From the total of 14 CCDs installed in the experiment 2 were disconnected due to 
issues that appeared at the beginning of the operation of the experiment. 
Of the 12 remaining detectors, we selected 8 that have shown good performance in terms of noise, charge transfer efficiency, and long-term stability. 

After removing edge effects, the effective size of each CCD is 4093 $\times$ 4111 pixels, giving a total mass of 47.6~g for the array of 8 CCDs. 
We remove from the analysis the columns that have an excess of hot pixels in comparison with the rest of the sensor. Hot columns detected on any image are eliminated from the analysis of the complete data set. This is done to ensure that we use the same parts of the detector in both reactor-on and reactor-off data sets. 

The total accumulated exposure of data considered in this work, corresponding to 8 CCDs operating during the two seasons, is 3.7~kg-days:
2.1~kg-days  
taken with the reactor on and 1.6~kg-days 
with the reactor off. 

\subsection{Low-energy event selection}
\label{eventselection}

As discussed in section~\ref{cataloggen}, events are selected to contain energy above a threshold ($Q_{\rm th}$) for the seed, which is set at 10~e$^{-}$, corresponding to roughly four times the standard deviation of the RN: 
$Q_{\rm th} = 10~$e$^{-} = 0.037$~keV. 
Some pixel fluctuations from on-chip noise sources (section~\ref{RNDC})  could be large enough to produce fake events that resemble low-energy neutrino events given this threshold. 
A statistical test is used to separate neutrino-like events from spurious ones from on-chip noise sources at low energies. 
The test is based on the likelihood of the pixel values of an event to follow the probability density function of the Gaussian readout noise. 
The log-likelihood $L_j$ is calculated from the $N$ pixels of the $j$-th event as:
\begin{equation}
    L_j(P_1,...,P_N | \sigma) = \sum^{N}_{i=1}(-1)^{n_i}\left(\frac{P_i^2}{2\sigma^2}+\log(\sqrt{2\pi}\sigma)\right)\,,
\end{equation}
where $P_i$ is the value of pixel $i$ (here we consider only level 0 and level 1 pixels)
and $n_i=1$ (0) if $P_i \geq 0$ ($P_i <0$). It should be noted that $n_i=1$ corresponds to a Gaussian probability distribution and $n_i=0$ is included to maximize the power of the statistical discriminator for negative fluctuations in the pixel value.
 The log-likelihood selection criterion
 is chosen to maintain contributions from on-chip noise much below the background radiation contribution (dominated by Compton scattering). 
 Images consisting purely of on-chip noise (DC and RN) were simulated to evaluate their distribution in $L_j$ and the on-chip noise contribution to the low-energy spectrum in the experiment.
 
 \begin{figure}[htb]
\centering
\includegraphics[scale=0.45]{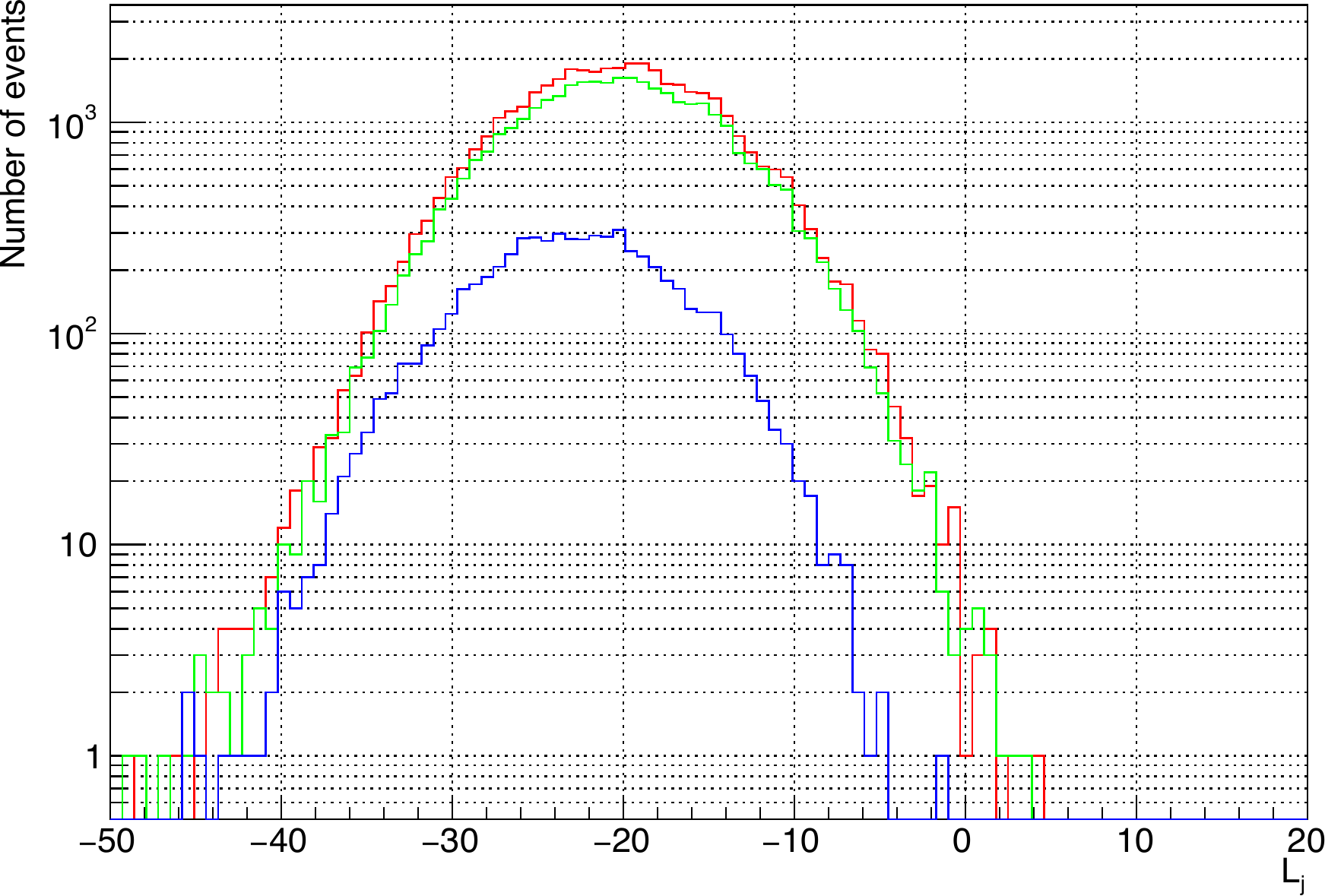} 
\caption{$L_j$ distribution from simulated noise 
events for three combinations of DC and RN representative of the selected data.} 
\label{fig:fake events likelihood distribution}
\end{figure}

 Fig.~\ref{fig:fake events likelihood distribution} shows the $L_j$ distribution of simulated events for three different conditions of DC and RN in our sensors, which are 
 representative of the values obtained during the selected periods (see sec.~\ref{RNDC}). 
 Each condition was evaluated over a group of 1500 images with similar size as the data of the experiment, equivalent to an exposure of 1.125 kg-days.  
 All the events with pixel seeds above $Q_{\rm th}$ are extracted and those with energy above 0.075 keV are evaluated by the likelihood and incorporated in the plot. 
 The red histogram (upper curve in this figure) represents the most extreme DC and RN condition for our sensors ($\lambda \simeq 0.25$~e$^-$/pix/hr and $\sigma \simeq 2~$~e$^-$),
  which gives the highest rate of fake events above threshold. 
%
 To prevent this systematic error from impacting
 the analysis, a cut of $L_j<-45$ was chosen, which gives a 
 a negligible contribution (less than two orders of magnitude) compared to the expected electromagnetic background contribution after fiducial cuts in the same energy range.

\subsection{Efficiency for CE$\nu$NS events}
\label{efficiencySec} 

\begin{figure}[htb]
\centering
\includegraphics[scale=0.455]{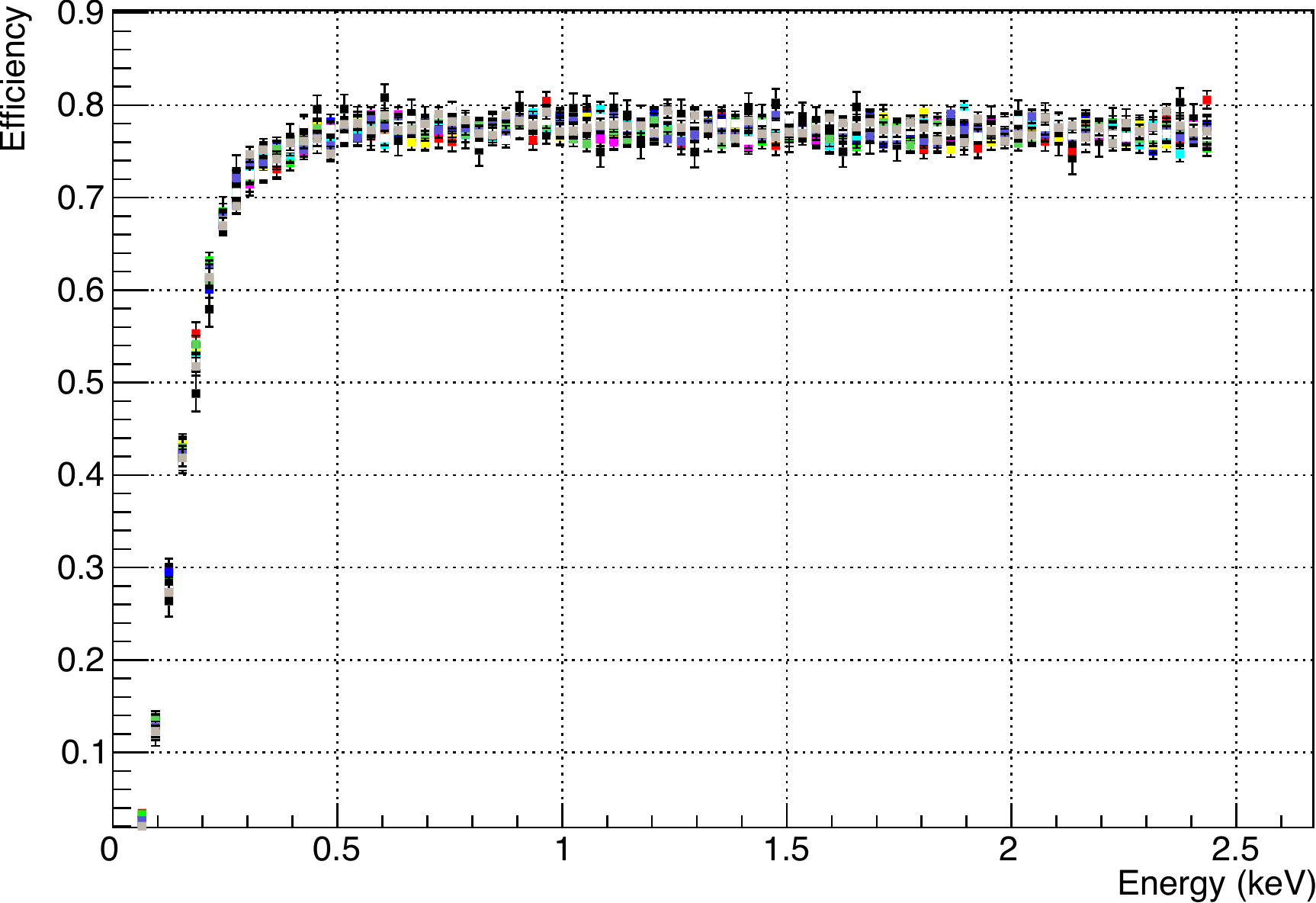} 
\caption{Reconstructed efficiency of neutrino-like events simulated in the vertical overscan region of one sensor. Each color corresponds to a different sub-run in the detector covering all data used for the analysis in the paper.}
\label{fig:reco_efficiency_with_time}
\end{figure}

The conservative selection cut $L_j<-45$ is used to separate neutrino-like events from the noise-like ones for all the sensors. 
%
To evaluate the detection efficiency at low energies, while ensuring that the small variations in the noise 
and other statistical fluctuations do not impact 
the reconstructed number of neutrino events in all data sets, simulated neutrino-like events are added in the vertical overscan of the output images of the experiment. 
Since this region has a very short exposure of about 15 minutes per image, they have almost no background events and the contribution of small RN fluctuations can be easily evaluated. 
The simulated neutrino events are then reconstructed using the same processing tools and their reconstruction efficiency is evaluated for each data set of every run. 
Neutrino-like simulated events with energies up to 2.5 keV were added and their reconstruction efficiency is shown in Fig.~\ref{fig:reco_efficiency_with_time} for all the sub-runs used in this analysis for one representative CCD. 
The fluctuations of the sub-run efficiencies at low energies (below 0.5 keV), where the $L_j$ calculation is more sensitive to small RN variations, are the same as for higher energies and all lie within the statistical 
uncertainty of the measurement. This shows that the different noise conditions do not impact the efficiency.

\begin{figure}[htb]
\centering
\includegraphics[scale=0.455]{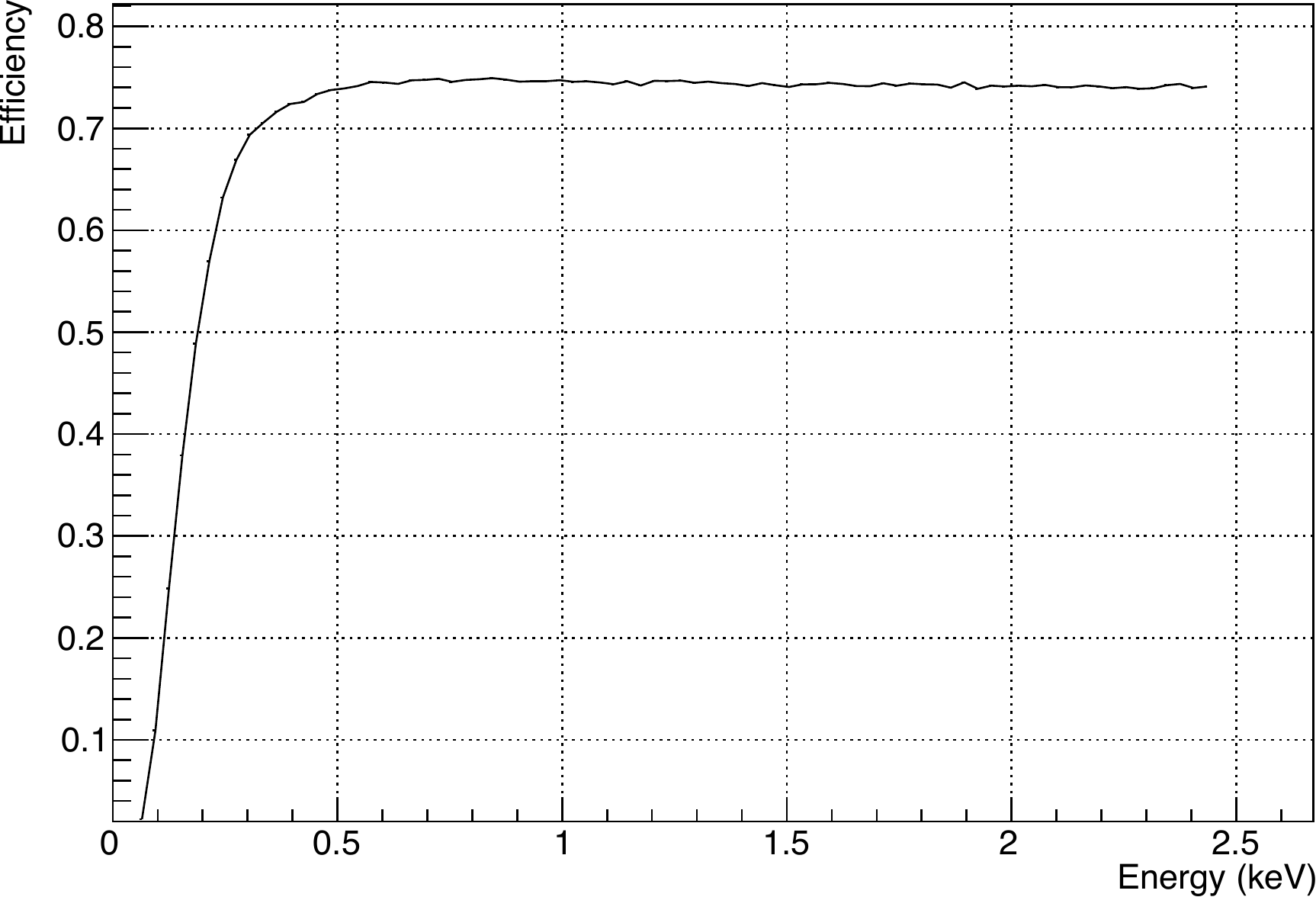}
\caption{Overall efficiency of neutrino event reconstruction.}
\label{fig: efficiency}
\end{figure}

To obtain an overall reconstruction efficiency for neutrinos for all CCDs and periods, we add neutrino-like events
on the active region of the sensor, which accounts for the reduction in the total exposure by overlapping with higher-energy events. 
Only data sets without expected neutrino events (reactor off periods) were used, to avoid any efficiency reduction due to the neutrino signal.
Neutrinos are simulated with uniform probability in the active volume of the sensor, with shape determined by the calibration in section~\ref{sec: size calibration} and with a uniform distribution in energy up to 2.5 keV. 
A set of 1000 events are simulated per CCD image. This number was chosen to provide a large enough sample to evaluate the efficiency with a low uncertainty, without having a significant impact on the total occupancy. 
The images with simulated neutrino events are 
processed using the standard chain and the selection rules described above are applied. The measured efficiency for each sensor is then weighted by the exposure, yielding 
the overall efficiency curve presented in Fig.~\ref{fig: efficiency}. 

\section{Search for the standard model CE$\nu$NS signal}
\label{CENNSsearch}

\begin{figure}[htb]
\centering
    \includegraphics[scale=0.49]{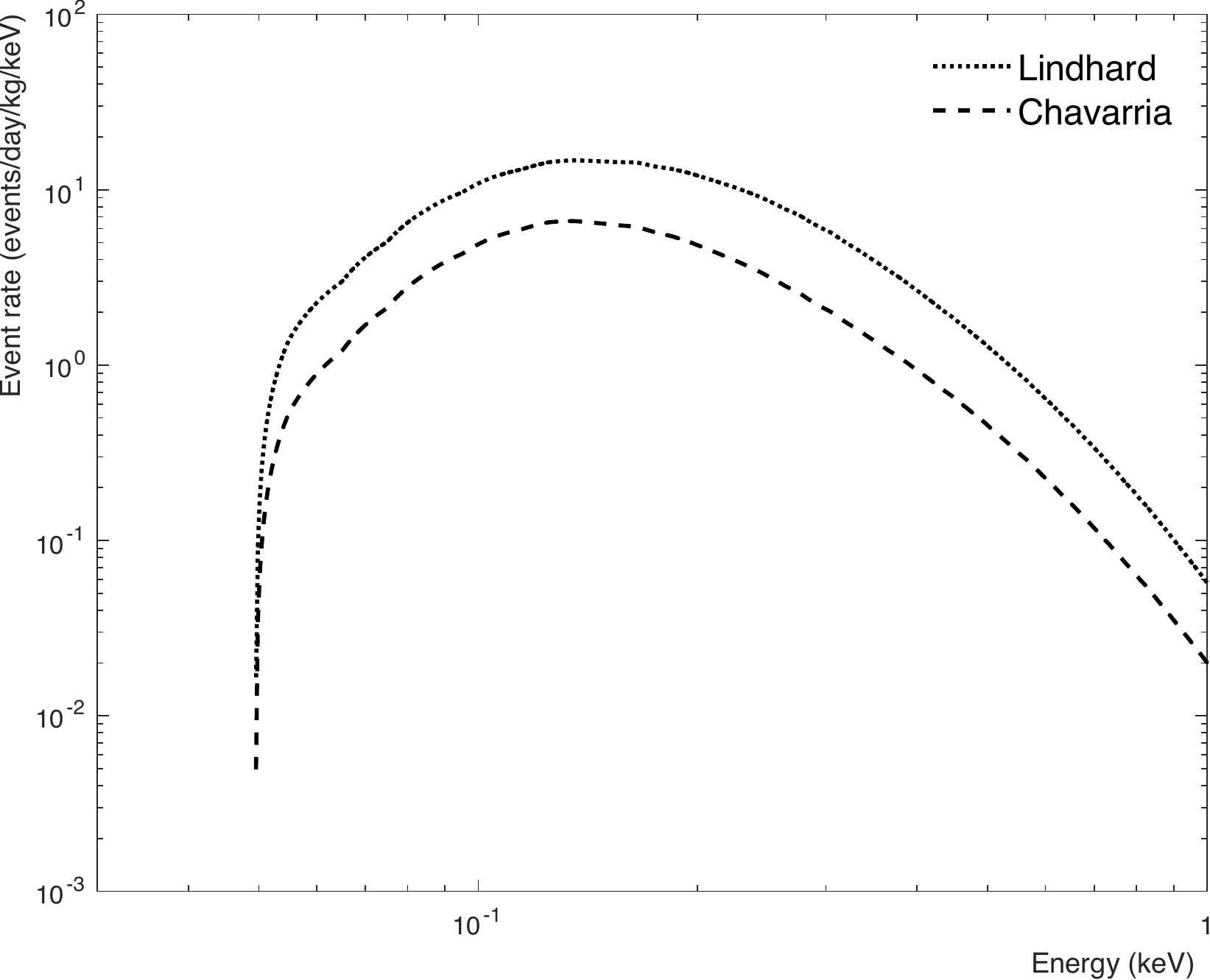} 
\caption{Observable neutrino recoil spectrum in the CONNIE detector array using two versions of the quenching factor measured from Lindhard et al.~\cite{lindhard} (dotted line) and Chavarria et al.~\cite{Chavarria:2016xsi} (dashed line).}
\label{fig: observable experiment}
\end{figure}

To obtain the expected rate of CE$\nu$NS events in the detector array we follow the prescription in~\cite{2015PhRvD..91g2001F}, using the overall efficiency discussed in the previous section. 
For the CE$\nu$NS process, the relevant energy is the silicon recoil energy. However, the energy calibration discussed in Sec.~\ref{calibsec} is based on the ionization produced by recoiling electrons from the photo absorption of x-rays of known energy. Therefore, all quoted energies are in electron-equivalent units.
The conversion from electron-equivalent to silicon recoil energy is given by the so-called quenching factor.
Here we employ a recent measurement of this
factor for nuclear recoils in CCDs from \cite{Chavarria:2016xsi} to compute the expected event rate from CE$\nu$NS.
For comparison to previous work \cite{connie:2016}, the expected event rate is also calculated using the quenching model from Lindhard \cite{lindhard}. 
The expected event rate is shown in Fig.~\ref{fig: observable experiment}, in observable (i.e., electron-equivalent) energy, for the two quenching factors.
The energy resolution for the detectors discussed in section~\ref{sec:eresolution} has been included in this calculation, smearing the measured energy of the simulated nuclear recoils.  
The recoil spectrum is convolved with a Gaussian resolution, where the width of the Gaussian varies according to the model of Eq.~(\ref{eq:energyResolution}). 
As a result of this smearing 
a fraction of low-energy events gets promoted to higher observable energies. 

\begin{figure}[htb]
\centering
    \includegraphics[scale=0.495]{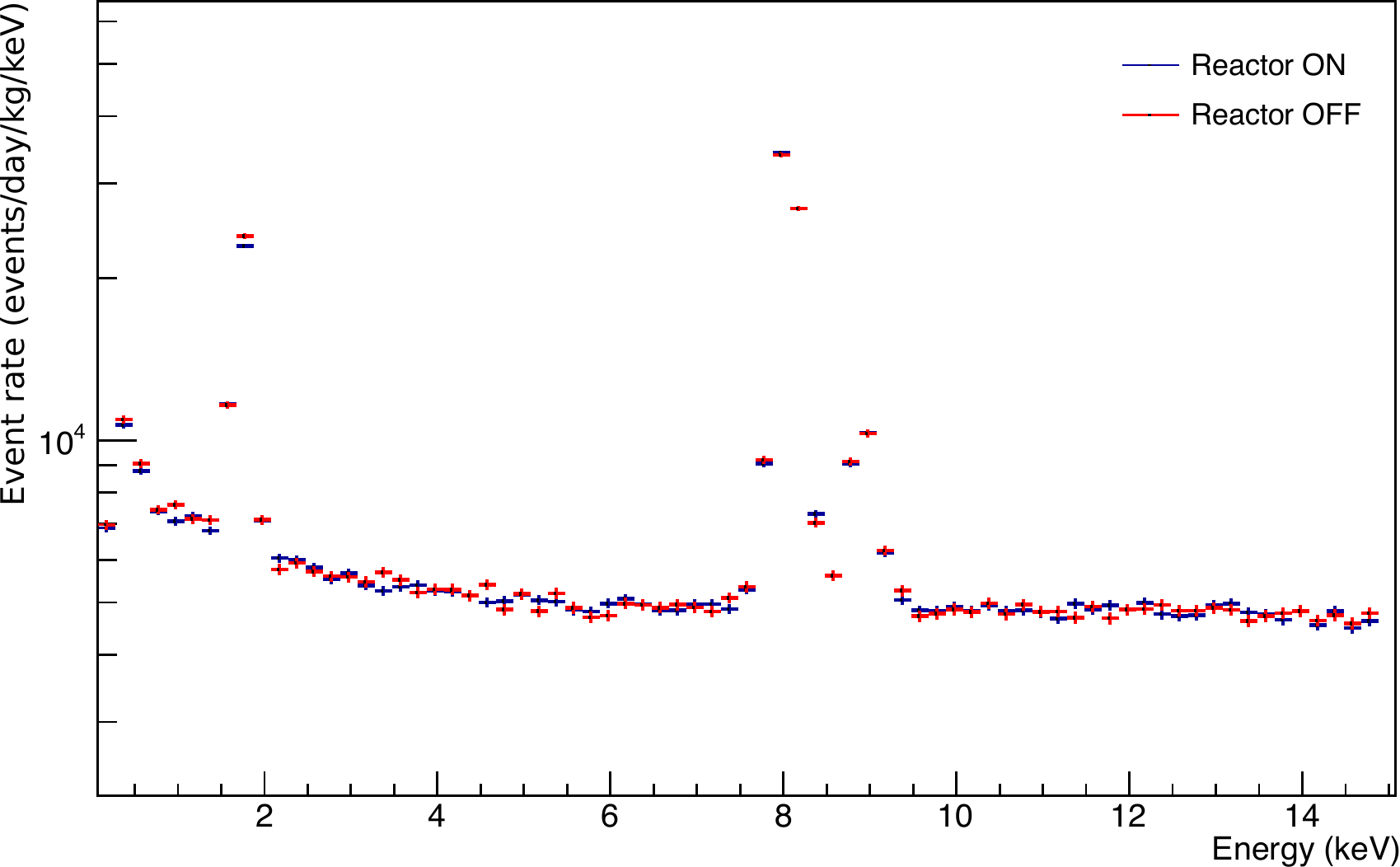} 
\caption{Energy spectrum for reactor-on and reactor-off data.}
\label{fig:reactor on and off data}
\end{figure}

To search for a CE$\nu$NS signal, the selection criteria discussed in section~\ref{eventselection} are applied to the data with the reactor on and off periods.
Fig.~\ref{fig:reactor on and off data} shows the observed spectrum for both periods at energies below 15~keV. 
The data for each CCD in the detector array are weighted by their exposure mass and included in this spectrum. 
The x-ray fluorescence lines for silicon 
in the sensor active volume and copper 
surrounding the sensors are clearly observed.

\begin{figure}[htb]
\centering
    \includegraphics[scale=0.49]{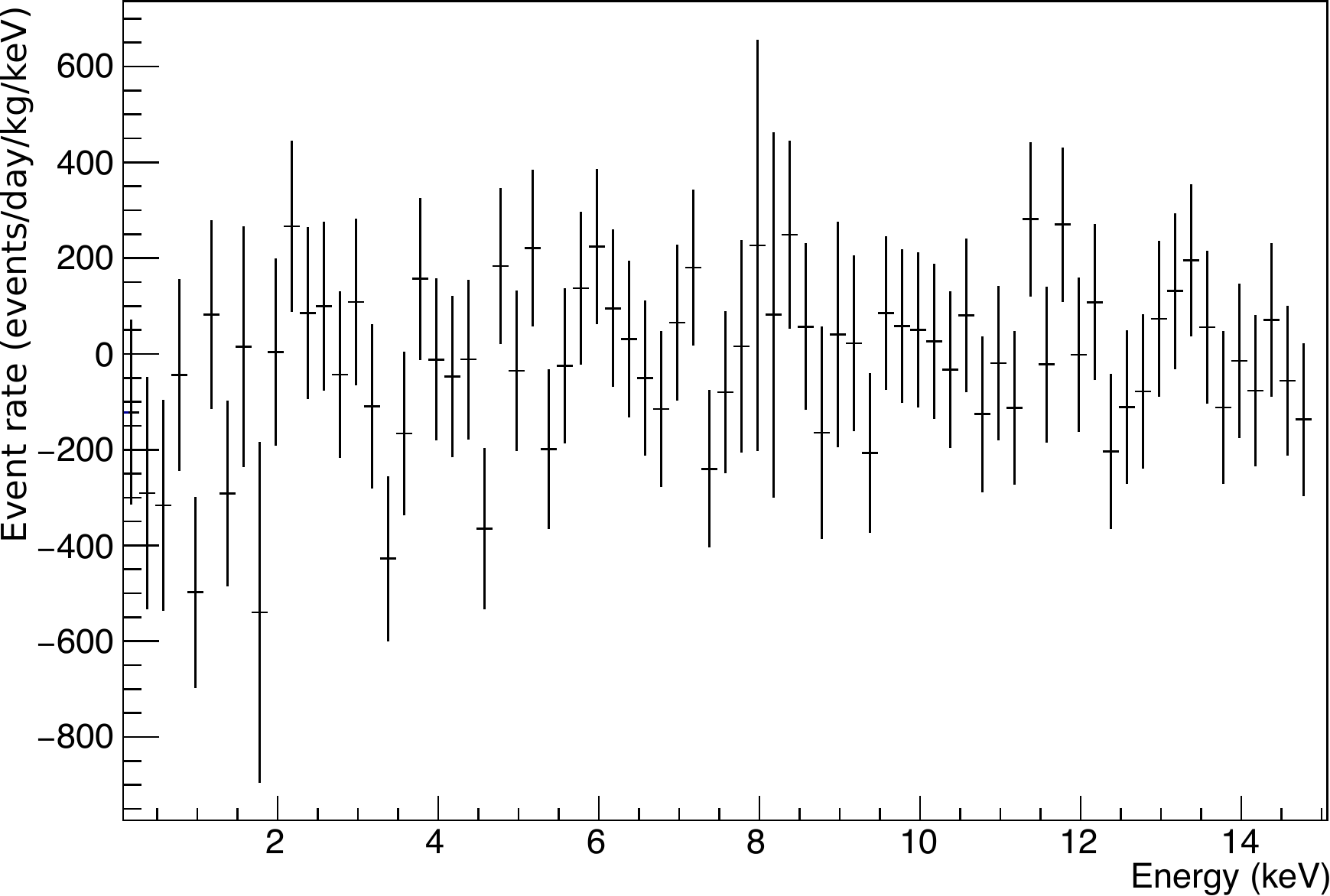}
\caption{Energy spectrum difference of reactor-on minus reactor-off data.}
\label{fig:reactor on minus off data 15keV}
\end{figure}

The reactor-off spectrum is subtracted from the reactor-on one for each sensor in the detector array and the results binned in energy are weighted by the sensor exposure mass and combined in Fig.~\ref{fig:reactor on minus off data 15keV}. 
The error bars in this figure reflect the statistical uncertainty in the binned spectrum subtraction. 

\begin{figure}[htb]
\centering
    \includegraphics[scale=0.5]{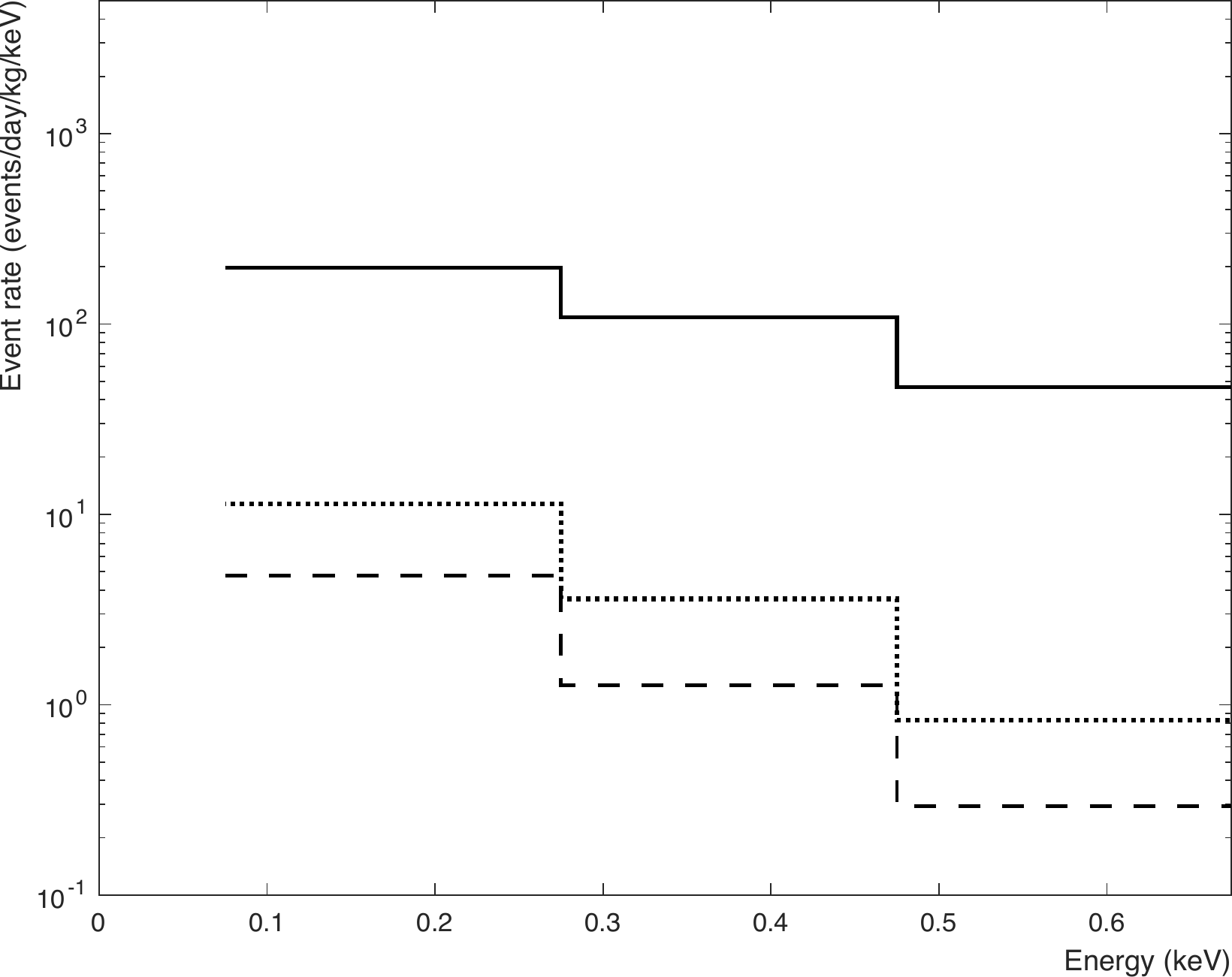} 
\caption{CE$\nu$NS event rate: 95\% confidence level limit from the reactor on - off measurement (solid line) and neutrino signal expected from the Lindhard~\cite{lindhard} (dotted line) and Chavarria~\cite{Chavarria:2016xsi} (dashed line) quenching factors.}
\label{fig:95CL limit}
\end{figure}

There is no significant excess of events in the reactor-on minus reactor-off subtraction. The maximum 
excess consistent with the data at 95\% confidence level (CL) is shown in Fig.~\ref{fig:95CL limit} and Table~\ref{Table:CENNSlimit}. 
This limit is compared to the expected CE$\nu$NS event rate using the quenching factor measured from Chavarria~\cite{Chavarria:2016xsi} and the Lindhard~\cite{lindhard} models. 
The results show that the 95\% CL limit established by this work is a factor of $\sim$40 above the prediction from the SM for deposited energies about 0.1~keV, 
or recoil energies of 1~keV. 

\begin{table}[htb]
\centering
\begin{tabular}{ |c|c|c|c| } 
\hline
Energy  & CE$\nu$NS-rate & CE$\nu$NS-rate & 95\% C. L.\\
 range (keV)  & 
   Lindhard  &  
   Chavarria & from data\\\hline
 0.075--0.275 & 11.4 & 4.8 & 197\\
 0.275--0.475 & 3.6 & 1.3 & 109\\
 0.475--0.675 & 0.8 & 0.3 & 47\\


 \hline
\end{tabular}
\caption{Expected rate from CE$\nu$NS, in events/day/kg/keV, assuming quenching factors from Lindhard~\cite{lindhard} and Chavarria~\cite{Chavarria:2016xsi} together with the 95\% CL limit from the data presented in this paper.}
\label{Table:CENNSlimit}
\end{table}

\section{Discussion and concluding remarks}

The CONNIE experiment is operated remotely at the Angra 2 nuclear power plant.
During 2017 and 2018 an operating efficiency of more than  95\% was achieved, 
thanks to a monitoring, alarms and interlock system developed to record and report the status of all the critical values of the experiment. 
The CONNIE results demonstrate the operation of low-threshold detectors next to a commercial power plant to search for CE$\nu$NS while maintaining good control of the reactor related background. 
The capability to monitor the stability of the environmental radiation background is also demonstrated thanks to the excellent energy resolution 
of the sensors. 

The results presented here constitute the first search for CE$\nu$NS at a nuclear reactor reaching recoil energies down to 1~keV (0.1~keV electron-equivalent). 
This measurement was made possible thanks to the development of a detector based on thick fully depleted low-threshold CCDs, specifically designed for this purpose. 
Low-threshold CCDs open a new window into the low-energy neutrino physics sector, probing for physics beyond the SM \cite{Okun:1986hi,Nelson:2007yq,Pospelov:2011ha, Essig:2013lka,Ilten:2018crw}. 
The threshold explored by CONNIE is one order of magnitude lower than the threshold of 20~keV used for the first detection of CE$\nu$NS by  the COHERENT experiment \cite{Coherent}. 

The data used in this paper allowed us to place an upper limit on the CE$\nu$NS event rate at about 40 times above the expectation from the SM. Due to the CONNIE low-energy threshold, this upper bound allows us to impose constraints on some NSI models, which can be competitive with the bounds from the COHERENT detection. For example, models with a light mediator~\cite{Harnik:2012ni}, which may increase the rate of events at the lowest energies by orders of magnitude, can be strongly constrained by the CONNIE data.

In this paper, rather than deriving the implications on specific models, we choose to present our results in a model independent way, in terms of the upper bounds on the event rates, as in Table~\ref{Table:CENNSlimit}. We expect them to be useful for other groups to investigate the constraints on different models and to compare them with the results and expectations from other experiments.

The sensitivity to the SM CE$\nu$NS obtained in this measurement is somewhat lower than the expectation from the forecast presented in reference \cite{2015PhRvD..91g2001F}. 
There are three reasons for the reduced sensitivity.  First, the updated measurements for the quenching factor in~\cite{Chavarria:2016xsi} reduce significantly the expected signal compared to the Lindhard model \cite{lindhard}. 
The CONNIE collaboration is working to reduce the uncertainty in the quenching factor, in collaboration with other teams using silicon targets for the detection of nuclear recoils \cite{damic:2016,Crisler:2018gci, Abramoff:2019dfb,superCDMS}. 
The second is the lower detection efficiency than the estimations used for the forecast in~\cite{2015PhRvD..91g2001F}. 
We expect to recover most of the efficiency by upgrading the experiment with the recently demonstrated skipper-CCD sensors~\cite{Tiffenberg:2017aac}. Finally, the low-energy background measured in the CONNIE experiment is about a factor of 10 higher than the estimations in the forecast. 

The CONNIE detector array was designed to have a geometry appropriate for track and shower reconstruction as an additional tool to identify background events, however these capabilities were not exploited for the analysis presented here. 
We expect to make use of the shower reconstruction capabilities of the detector in future work, extending the sensitivity of the experiment.

The analysis discussed here for the CONNIE data is based on a reactor-on minus reactor-off subtraction. 
This model-independent analysis is strongly limited by the statistics of the reactor-off data, equivalent to less than 10\% of the total data that is possible to collect. 
A model-dependent analysis using the spectral details of signal and background based on a full simulation of the detector at low energies will increase the sensitivity to the SM CE$\nu$NS signal and is planned for future work.


\acknowledgments

We thank Eletrobras Eletronuclear for access to the Angra 2 reactor site and for the support of their personnel, in particular Ilson Soares and Gustavo Coelho, to the CONNIE activities. We thank the Silicon Detector Facility team at Fermi National Accelerator Laboratory for being the host lab for the assembly and testing of the detectors components used in the CONNIE experiment.
We acknowledge Marcelo Giovani for his IT support.
The Mexican group acknowledges Ing. Mauricio Mart\'inez for his technical assistance. 
CB and MM acknowledge the hospitality of Fermilab, where part of this work was done. 
We acknowledge the support from the former Brazilian Ministry for Science, Technology, and Innovation (currently MCTIC), the PCI program at CBPF and the Brazilian funding agencies FAPERJ (grants E-26/110.145/2013, E-26/210.151/2016), CNPq, and FINEP (RENAFAE grant 01.10.0462.00); and M\'exico's CONACYT (grant No. 240666) and DGAPA-UNAM (PAPIIT grant IN108917).
This work made use of the CHE cluster, managed and funded by COSMO/CBPF/MCTI, with financial support from FINEP and FAPERJ, and operating at the Javier Magnin Computing Center/CBPF. 

\bibliography{CONNIEbib}

\end{document}